# Design and Performance Analysis of Dual and Multi-hop Diffusive Molecular Communication Systems

Neeraj Varshney, *Student Member, IEEE*, Adarsh Patel, *Student Member, IEEE*, Aditya K. Jagannatham, *Member, IEEE*, and Pramod K. Varshney, *Fellow, IEEE*

*Abstract*—This work presents a comprehensive performance analysis of diffusion based direct, dual-hop, and multi-hop molecular communication systems with Brownian motion and drift in the presence of various distortions such as inter-symbol interference (ISI), multi-source interference (MSI), and counting errors. Optimal decision rules are derived employing the likelihood ratio tests (LRTs) for symbol detection at each of the cooperative as well as the destination nanomachines. Further, closed-form expressions are also derived for the probabilities of detection, false alarm at the individual cooperative, destination nanomachines, as well as the overall end-to-end probability of error for source-destination communication. The results also characterize the impact of detection performance of the intermediate cooperative nanomachine(s) on the end-to-end performance of dual/multi hop diffusive molecular communication systems. In addition, capacity expressions are also derived for direct, dual-hop, and multi-hop molecular communication scenarios. Simulation results are presented to corroborate the theoretical results derived and also, to yield insights into system performance.

## I. INTRODUCTION

Nanoscale molecular communication has garnered significant interest in recent times towards addressing challenging problems in biomedical, industrial, and surveillance scenarios [1], [2]. This has lead to the development of novel applications such as efficient drug delivery and human body monitoring, using communicating nano-robots [3]–[5]. In contrast to active molecular communication (AMC)-based systems such as molecular motors and protein filaments, the molecules in diffusion-based passive molecular communication (PMC) systems propagate via Brownian motion in a fluidic medium without requiring additional infrastructure or external energy [1], [2]. Several research efforts [6]–[17] have been devoted to exploring various aspects such as developing channel models, estimating the channel as well as analyzing the performance of diffusion based molecular systems. It has been shown in [18] that the molecular concentration decays inversely as the cube of the distance between the transmitter and receiver nanomachines, which severely limits the performance of such systems. Relay-assisted cooperative communication has been shown to successfully overcome this impediment by significantly enhancing the communication range. This leads to a substantial improvement in the end-to-end reliability of communication, thus making it a very promising technology for such systems [19]. However, the decoding accuracy at the destination in relay-assisted molecular communication depends critically on the detection performance of the intermediate cooperative nanomachine(s) that act as relays. The end-to-end performance can further deteriorate due to other degrading effects such as inter-symbol interference (ISI), multi-source interference (MSI), and counting errors at the cooperative as well as destination nanomachines [20]. ISI at the receiving nanomachine arises due to Brownian motion, which results in the molecules emitted by the transmitting nanomachine at the beginning of given time slot arriving stochastically in subsequent time-slots. On the other hand, MSI arises from the transmissions of other sources using the same type of molecules. Some works in the existing literature [21]–[29] propose techniques for receiver design and detection for direct source-destination diffusive molecular communication. However, to the best of our knowledge, none of the works in the existing literature have analyzed the impact of detection performance of the intermediate cooperative nanomachine(s) on the end-to-end performance of the relay-assisted dual and multi-hop molecular communication in the presence of ISI, MSI, and counting errors and is, one of the central aims of the work presented in this paper. Next, we present a detailed overview and comparative survey of related works in the existing literature on relay-assisted molecular communication.

### A. Related Work

Some works in the existing literature have analyzed the performance of single relay-assisted molecular communication systems. In [30], a sense-and-forward relaying strategy has been proposed for diffusion based molecular communication between two nodes formed from synthetic bacteria. The analysis is extended in [31] to decode-and-forward (DF) relaying of $M$-ary information symbols where it was demonstrated that optimal combining of the direct and relayed outputs improves communication reliability. Authors in [32] derived an expression to characterize the average error probability for a two-hop DF molecular network and subsequently proposed mitigation techniques for the self-intereference that arises at the relay due

Neeraj Varshney, Adarsh Patel and Aditya K. Jagannatham are with the Department of Electrical Engineering, Indian Institute of Technology Kanpur, Kanpur UP 208016, India (e-mail:{neerajv; adarsh; adityaj}@iitk.ac.in).

Pramod K. Varshney is with the Department of Electrical Engineering & Computer Science, Syracuse University, Syracuse, NY 13244, USA (e-mail:varshney@syr.edu).

to the detection and emission of the same type of molecules. The analysis is further extended to amplify-and-forward (AF) relaying with fixed and variable amplification factors in [33]. The optimal amplification factor at the relay node to minimize the average error probability of the network is also derived therein based on an approximation of the same. The authors in [34] proposed an energy efficient scheme for the information molecule synthesis process employing a simultaneous molecular information and energy transfer (SMIET) relay. Performance analysis was presented for the resulting bit error probability and cost of synthesis. However, the above works [30]–[34] do not consider practical effects such as MSI, ISI and counting errors at any of the receiver nanomachines. The work in [35] analyzed the bit error rate (BER) performance of a dual-hop DF molecular communication system in which the time-dependent molecular concentrations are influenced by noise and ISI resulting from channel. However, the work in [35] does not consider the effect of MSI while analyzing the performance.

In the context of multi-hop communication, the design and analysis of repeater cells in Calcium junction channels was investigated in [36], where signal molecules released by the transmitter are amplified by intermediate repeaters in order to aid them reach the receiver. The authors in [32] extended the error probability analysis to multi-hop links in [37]. In [38], a diffusion-based multi-hop network between bacterial colonies was analyzed with several bacteria agents acting as a single node. Further, the use of bacteria and virus particles as information carriers in a multi-hop network was proposed in [39] and [40], respectively.

The work in [41] recently analyzed the error rate performance of a diffusion based DF dual-hop molecular communication system inside a blood vessel of a human body. The analysis therein formulates an optimization problem to find the optimal threshold which minimizes the bit error probability (BEP) at the destination nanomachine. However, analytical results are not presented therein for the same. None of the related studies in the existing literature consider the problem of optimal detection at the destination nanomachine using LRT for binary hypothesis testing in diffusion based dual-hop and multi-hop molecular communication systems inside the blood vessels of the human body. Moreover, the presence of practical aberrations such as the ISI, MSI, and counting errors and the impact of the detection performance of the intermediate cooperative nanomachines on the end-to-end performance of relay-assisted dual and multi-hop molecular communication has not been studied in the literature.

### B. Contributions

The main contributions of the paper can be logically organized into three parts as described.

1) Initially, the LRT-based optimal decision rule is determined at the destination nanomachine in the presence of ISI, MSI, and counting errors in a diffusion based direct molecular communication system with drift. Further, closed-form expressions are derived for the probabilities of detection and false alarm are derived to analytically characterize the detection performance at the destination nanomachine. These results are subsequently used to determine the probability of error and the channel capacity for molecular communication between the source and destination nanomachines.

2) In the second part of the paper, the above results are extended to a dual-hop scenario with DF relaying in which the cooperative nanomachine decodes the symbol using the number of molecules received from the source nanomachine and forwards it to the destination nanomachine in the subsequent time-slot. In contrast to [41], closed-form expressions are presented for the optimal decision rule including optimal threshold, resulting detection performance, and capacity. Further, the probabilities of detection and false alarm are also derived at the intermediate cooperative nanomachines.

3) The Final part presents a general framework to analyze the performance of a multi-hop molecular communication scenario with $N \geq 2$ cooperative nanomachines. The optimal LRT based decision rule, probabilities of detection and false alarm are derived at each cooperative and destination nanomachine incorporating the performance of the intermediate cooperating nonomachines. The results are subsequently used to obtain analytical results for the end-to-end probability of error and capacity for the aforementioned multi-hop scenario.

### C. Organization

The rest of the paper is organized as follows. The system model for diffusion based direct communication between the source and destination nanomachines is presented in Section II-A. Comprehensive analyses of the probabilities of detection, and false alarm at the individual cooperative and destination nanomachines, along with the end-to-end probability of error and capacity are presented in Section II-B, Section II-C, and Section II-D respectively. This is followed by similar results for the dual-hop and multi-hop communication cases in Sections III and IV respectively. Simulation results for all the above scenarios are presented in Section V, followed by conclusions in Section VI.

## II. DIRECT SOURCE-DESTINATION COMMUNICATION

### A. System Model

Consider a diffusion based molecular communication system inside a blood vessel in which the source nanomachine communicates with the destination nanomachine directly in a fluid medium with positive drift. In this system, the source nanomachine is assumed to be a point source with the destination nanomachine located at a distance of $d_{sd}$ from the source nanomachine. Further, similar to [8], [20], [41] and the references therein, both nanomachines are assumed to be synchronized in time and stationary. The channel is divided into time slots of duration $\tau$ as shown in Fig. 1, where the $j$th slot is defined as the time period $[(j-1)\tau, j\tau]$ with $j \in \{1, 2, \cdots\}$. At the beginning of each time slot, the source nanomachine emits either the same type of molecules in the propagation medium for information symbol 1 generated with a prior probability $\beta$ or remains silent for information symbol 0. Let $\mathcal{Q}_0[j]$ denote the number of type[1] 0 molecules released by the source nanomachine to transmit the information symbol $x[j] = 1$ at the beginning of the $j$th slot. The molecules released by the source nanomachine diffuse freely in the propagation medium and are assumed to be in Brownian motion with a positive drift in the direction of information transmission from the source to destination nanomachines. Similar to the works in [20], [41] and the references therein, this work also assumes that the transmitted molecules do not interfere or collide with each other. Moreover, once these molecules reach the destination nanomachine, they are assumed to be immediately absorbed by it and not propagate further in the medium. Finally, the destination nanomachine decodes the information symbol transmitted by the source nanomachine using the number of molecules received at the end of each time slot.

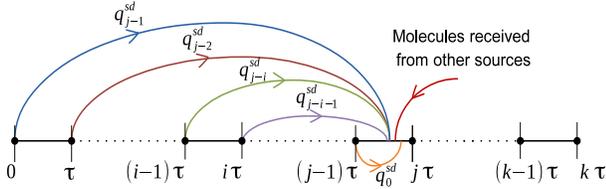

Fig. 1. Diffusion based molecular communication over time-slotted channel [20], where $[(j − 1)\tau, j\tau]$ is the current slot and $q_0^{sd}$ is the probability that a molecule reaches the destination nanomachine within the current slot.

The molecules emitted by the source nanomachine arrive stochastically at the destination nanomachine and degrade over time. This is owing to the fact that the life expectancy of a molecule decays with time. Let $q_{j-i}^{sd}$ denote the probability that a molecule transmitted in slot $i \in \{1, 2, \cdots, j\}$ arrives at the destination nanomachine in time slot $j$, which can be obtained as [8, Eq. (1)]

$$q_{j-i}^{sd} = \int_{(j-i)\tau}^{(j-i+1)\tau} f(t) \int_t^\infty g(u) du dt, \quad (1)$$

where $f(t)$ is the probability density function (PDF) of the first hitting time, i.e., the time required for a molecule to reach the destination nanomachine, and $g(u)$ denotes the PDF of the molecular life expectancy. The first hitting time $f(t)$ follows the inverse Gaussian distribution with PDF [43], [44]

$$f(t) = \sqrt{\frac{\lambda}{2\pi t^3}} \exp\left(-\frac{\lambda(t-\mu)^2}{2\mu^2 t}\right), t > 0, \quad (2)$$

where the mean $\mu$ and the shape parameter $\lambda$ are given as, $\mu = \frac{d_{sd}}{v}$ and $\lambda = \frac{d_{sd}^2}{2D}$ respectively, with $d_{sd}$ denoting the distance between the source and destination nanomachines, $v, D$ denoting the drift velocity and diffusion coefficient of the molecules respectively. Further, similar to [8], [20], the life expectancy is modeled as an exponential distribution $g(u) = \alpha \exp(-\alpha u), u > 0$, with mean $\frac{1}{\alpha}$, where $\alpha$ is referred to as the degradation parameter.

[1]Subscript 0 in $\mathcal{Q}_0[j]$ is simply used to denote a type of molecules considered for transmission by the source nanomachine. In the multi-hop scenario, we assume that each cooperative nanomachines sends different types of molecules. Therefore, we later used subscripts $1, 2, \cdots, N$ to denote $N$ different types of molecules transmitted by cooperative nanomachines $R_1, R_2, \cdots, R_N$ respectively. Further, it is worth mentioning that nanomachines such as eukaryotic cells [42] can be genetically modified to emit different types of molecules.

The number of molecules received at the destination nanomachine during the time slot $[(j-1)\tau, j\tau]$ can be expressed as

$$R_{sd}[j] = S_{sd}[j] + \mathcal{I}_{sd}[j] + N_{sd}[j] + C_{sd}[j]. \quad (3)$$

The quantity $S_{sd}[j]$ is the number of molecules received in the current slot $[(j-1)\tau, j\tau]$ and follows a binomial distribution [45], [46] with parameters $\mathcal{Q}_0[j]x[j]$ and $q_0^{sd}$, i.e., $Binomial(\mathcal{Q}_0[j]x[j], q_0^{sd})$, where $x[j] \in \{0, 1\}$ is the symbol transmitted by the source nanomachine in the $j$th time slot. The quantity $N_{sd}[j]$ denotes MSI, i.e., noise arising due to molecules received from other sources, which can be modeled as a Gaussian distributed random variable with mean $\mu_o$ and variance $\sigma_o^2$ under the assumption that the number of interfering sources is sufficiently large [47]. Also, note that the noise $N_{sd}[j]$ and the number of molecules $S_{sd}[j], \mathcal{I}_{sd}[j]$ received from the intended transmitter nanomachine are independent [20]. The term $C_{sd}[j]$ denotes the error in counting the molecules at the destination nanomachine, also termed as the "counting error", and can be modeled as a Gaussian distributed random variable with zero mean and variance $\sigma_c^2[j]$ that depends on the average number of molecules received as, $\sigma_c^2[j] = \mathbb{E}\{R_{sd}[j]\}$ [20], [48]. The quantity $\mathcal{I}_{sd}[j]$ is the ISI arising in slot $j$ due to transmissions in the previous $j-1$ slots and is given as

$$\mathcal{I}_{sd}[j] = I_{sd}[1] + I_{sd}[2] + \cdots + I_{sd}[j-1], \quad (4)$$

where $I_{sd}[i] \sim Binomial(\mathcal{Q}_0[j-i]x[j-i], q_i^{sd}), 1 \leq i \leq j-1$, denotes the number of stray molecules received from transmission in the $(j-i)$th slot. If the number of molecules released by the source nanomachine is sufficiently large, the binomial distribution for $S_{sd}[j]$ can be approximated by the Gaussian distribution[2] with mean $\mu_{sd}[j] = \mathcal{Q}_0[j]x[j]q_0^{sd}$ and variance $\sigma_{sd}^2[j] = \mathcal{Q}_0[j]x[j]q_0^{sd}(1 - q_0^{sd})$, i.e., $S_{sd}[j] \sim \mathcal{N}(\mathcal{Q}_0[j]x[j]q_0^{sd}, \mathcal{Q}_0[j]x[j]q_0^{sd}(1 - q_0^{sd}))$ [49]. Similarly, the binomial distribution of $I_{sd}[i], 1 \leq i \leq j-1$ can be approximated as

$$I_{sd}[1] \sim \mathcal{N}(\mu_I[1] = \mathcal{Q}_0[j-1]x[j-1]q_1^{sd},$$
$$\sigma_I^2[1] = \mathcal{Q}_0[j-1]x[j-1]q_1^{sd}(1 - q_1^{sd})),$$
$$\vdots$$
$$I_{sd}[j-1] \sim \mathcal{N}(\mu_I[j-1] = \mathcal{Q}_0[1]x[1]q_{j-1}^{sd},$$
$$\sigma_I^2[j-1] = \mathcal{Q}_0[1]x[1]q_{j-1}^{sd}(1 - q_{j-1}^{sd})).$$

Further note that $S_{sd}[j]$ and $I_{sd}[i], i = 1, 2, \cdots, j-1$ are independent since the molecules transmitted in different time slots do not interfere with each other [20], [41].

[2]This approximation is reasonable when $\mathcal{Q}_0[j]q_0^{sd} > 5$ and $\mathcal{Q}_0[j](1 - q_0^{sd}) > 5$ [20].

## B. Detection Performance Analysis

Using the model in (3), the symbol detection problem at the destination nanomachine can be formulated as the binary hypothesis testing problem

$$\begin{aligned}\mathcal{H}_0 : R_{sd}[j] =& I_{sd}[1] + I_{sd}[2] + \cdots + I_{sd}[j-1] \\ &+ N_{sd}[j] + C_{sd}[j] \\ \mathcal{H}_1 : R_{sd}[j] =& S_{sd}[j] + I_{sd}[1] + I_{sd}[2] + \cdots + I_{sd}[j-1] \\ &+ N_{sd}[j] + C_{sd}[j],\end{aligned} \quad (5)$$

where the null and alternative hypotheses $\mathcal{H}_0$ and $\mathcal{H}_1$ correspond to the transmission of binary symbols 0 and 1 respectively during the $j$th time slot. The number of molecules received $R_{sd}[j]$ at the destination nanomachine corresponding to the individual hypotheses are distributed as

$$\begin{aligned}\mathcal{H}_0 : R_{sd}[j] &\sim \mathcal{N}(\mu_{sd,0}[j], \sigma^2_{sd,0}[j]) \\ \mathcal{H}_1 : R_{sd}[j] &\sim \mathcal{N}(\mu_{sd,1}[j], \sigma^2_{sd,1}[j]),\end{aligned} \quad (6)$$

where the mean $\mu_{sd,0}[j]$ and variance $\sigma^2_{sd,0}[j]$ under hypothesis $\mathcal{H}_0$ are derived as

$$\begin{aligned}\mu_{sd,0}[j] =& \mu_I[1] + \mu_I[2] + \cdots + \mu_I[j-1] + \mu_o \\ =& \beta \sum_{i=1}^{j-1} \mathcal{Q}_0[j-i] q_i^{sd} + \mu_o,\end{aligned} \quad (7)$$

$$\begin{aligned}\sigma^2_{sd,0}[j] =& \sigma^2_I[1] + \sigma^2_I[2] + \cdots + \sigma^2_I[j-1] + \sigma^2_o + \sigma^2_c[j] \\ =& \sum_{i=1}^{j-1} [\beta \mathcal{Q}_0[j-i] q_i^{sd}(1-q_i^{sd}) + \beta(1-\beta) \\ &\times (\mathcal{Q}_0[j-i] q_i^{sd})^2] + \sigma^2_o + \mu_{sd,0}[j],\end{aligned} \quad (8)$$

and the mean $\mu_{sd,1}[j]$ and variance $\sigma^2_{sd,1}[j]$ under hypothesis $\mathcal{H}_1$ are determined as

$$\begin{aligned}\mu_{sd,1}[j] =& \mu_{sd}[j] + \mu_I[1] + \mu_I[2] + \cdots + \mu_I[j-1] + \mu_o \\ =& \mathcal{Q}_0[j] q_0^{sd} + \beta \sum_{i=1}^{j-1} \mathcal{Q}_0[j-i] q_i^{sd} + \mu_o,\end{aligned} \quad (9)$$

$$\begin{aligned}\sigma^2_{sd,1}[j] =& \sigma^2_{sd}[j] + \sigma^2_I[1] + \sigma^2_I[2] + \cdots + \sigma^2_I[j-1] + \sigma^2_o + \sigma^2_c[j] \\ =& \mathcal{Q}_0[j] q_0^{sd}(1 - q_0^{sd}) + \sum_{i=1}^{j-1}[\beta \mathcal{Q}_0[j-i] q_i^{sd}(1-q_i^{sd}) \\ &+ \beta(1-\beta)(\mathcal{Q}_0[j-i] q_i^{sd})^2] + \sigma^2_o + \mu_{sd,1}[j]. \end{aligned} \quad (10)$$

Using the conditional PDFs $p(R_{sd}[j]|\mathcal{H}_0)$ and $p(R_{sd}[j]|\mathcal{H}_1)$ given in (6), the following result presents the LRT-based decision rule at the destination nanomachine for symbol detection.

*Theorem 1:* The LRT-based optimal decision rule for deciding 0 and 1 at the destination nanomachine corresponding to source nanomachine transmission during the $j$th time slot is obtained as

$$T(R_{sd}[j]) = R_{sd}[j] \underset{\mathcal{H}_0}{\overset{\mathcal{H}_1}{\gtrless}} \gamma'_{sd}[j], \quad (11)$$

where $\gamma'_{sd}[j]$ is the optimal decision threshold and is given as

$$\gamma'_{sd}[j] = \sqrt{\gamma_{sd}[j]} - \alpha_{sd}[j], \quad (12)$$

where the quantities $\gamma_{sd}[j]$ and $\alpha_{sd}[j]$ are defined as

$$\alpha_{sd}[j] = \frac{\mu_{sd,1}[j]\sigma^2_{sd,0}[j] - \mu_{sd,0}[j]\sigma^2_{sd,1}[j]}{\sigma^2_{sd,1}[j] - \sigma^2_{sd,0}[j]}, \quad (13)$$

$$\begin{aligned}\gamma_{sd}[j] =& \frac{2\sigma^2_{sd,1}[j]\sigma^2_{sd,0}[j]}{\sigma^2_{sd,1}[j] - \sigma^2_{sd,0}[j]} \ln\left[\frac{(1-\beta)}{\beta} \sqrt{\frac{\sigma^2_{sd,1}[j]}{\sigma^2_{sd,0}[j]}}\right] + (\alpha_{sd}[j])^2 \\ &+ \frac{\mu^2_{sd,1}[j]\sigma^2_{sd,0}[j] - \mu^2_{sd,0}[j]\sigma^2_{sd,1}[j]}{\sigma^2_{sd,1}[j] - \sigma^2_{sd,0}[j]}.\end{aligned} \quad (14)$$

*Proof:* The optimal log likelihood ratio test (LLRT) at the destination nanomachine can be obtained as

$$\Lambda(R_{sd}[j]) = \ln\left[\frac{p(R_{sd}[j]|\mathcal{H}_1)}{p(R_{sd}[j]|\mathcal{H}_0)}\right] \underset{\mathcal{H}_0}{\overset{\mathcal{H}_1}{\gtrless}} \ln\left[\frac{1-\beta}{\beta}\right]. \quad (15)$$

Substituting the Gaussian PDFs $p(R_{sd}[j]|\mathcal{H}_1)$ and $p(R_{sd}[j]|\mathcal{H}_0)$ from (6), the test statistic $\Lambda(R_{sd}[j])$ can be obtained as

$$\begin{aligned}\Lambda(R_{sd}[j]) =& \ln\left[\sqrt{\frac{\sigma^2_{sd,0}[j]}{\sigma^2_{sd,1}[j]}}\right] + \frac{1}{2\sigma^2_{sd,0}[j]\sigma^2_{sd,1}[j]} \\ &\times \underbrace{(R_{sd}[j]-\mu_{sd,0}[j])^2 \sigma^2_{sd,1}[j] - (R_{sd}[j]-\mu_{sd,1}[j])^2 \sigma^2_{sd,0}[j]}_{\triangleq f(R_{sd}[j])}. \end{aligned} \quad (16)$$

The expression for $f(R_{sd}[j])$ given above can be further simplified as

$$\begin{aligned}f(R_{sd}[j]) =& R^2_{sd}[j](\sigma^2_{sd,1}[j] - \sigma^2_{sd,0}[j]) \\ &+ 2R_{sd}[j](\mu_{sd,1}[j]\sigma^2_{sd,0}[j] - \mu_{sd,0}[j]\sigma^2_{sd,1}[j]) \\ &+ (\mu^2_{sd,0}[j]\sigma^2_{sd,1}[j] - \mu^2_{sd,1}[j]\sigma^2_{sd,0}[j]) \\ =& (\sigma^2_{sd,1}[j] - \sigma^2_{sd,0}[j])(R_{sd}[j] + \alpha_{sd}[j])^2 \\ &- \frac{(\mu_{sd,1}[j]\sigma^2_{sd,0}[j] - \mu_{sd,0}[j]\sigma^2_{sd,1}[j])^2}{\sigma^2_{sd,1}[j] - \sigma^2_{sd,0}[j]} \\ &+ (\mu^2_{sd,0}[j]\sigma^2_{sd,1}[j] - \mu^2_{sd,1}[j]\sigma^2_{sd,0}[j]),\end{aligned} \quad (17)$$

where $\alpha_{sd}[j]$ is defined in (13). Substituting the above equation for $f(R_{sd}[j])$ in (16) and subsequently grouping the terms independent of the received molecules $R_{sd}[j]$ with the threshold, the test reduces to

$$(R_{sd}[j] + \alpha_{sd}[j])^2 \underset{\mathcal{H}_0}{\overset{\mathcal{H}_1}{\gtrless}} \gamma_{sd}[j], \quad (18)$$

where $\gamma_{sd}[j]$ is defined in (14). Further, taking the square root of both sides with $\gamma_{sd}[j] \geq 0$, the above expression can be simplified to yield the optimal test given in (11). ∎

The detection performance at the destination nanomachine for the test in (11) is obtained next.

*Theorem 2:* The average probabilities of detection $(P_D)$ and false alarm $(P_{FA})$ at the destination nanomachine corresponding to the transmission by the source nanomachine in slots 1 to $k$ in the diffusion based molecular communication

nano-network are given as

$$P_D = \frac{1}{k}\sum_{j=1}^{k} P_D^d[j]$$
$$= \frac{1}{k}\sum_{j=1}^{k} Q\left(\frac{\gamma'_{sd}[j] - \mu_{sd,1}[j]}{\sigma_{sd,1}[j]}\right), \quad (19)$$

$$P_{FA} = \frac{1}{k}\sum_{j=1}^{k} P_{FA}^d[j]$$
$$= \frac{1}{k}\sum_{j=1}^{k} Q\left(\frac{\gamma'_{sd}[j] - \mu_{sd,0}[j]}{\sigma_{sd,0}[j]}\right), \quad (20)$$

where $P_D^d[j]$ and $P_{FA}^d[j]$ denote the probabilities of detection and false alarm, respectively, at the destination nanomachine in the $j$th time slot. The quantity $\gamma'_{sd}[j]$ is defined in (12) and the function $Q(\cdot)$ is the tail probability of the standard normal random variable.

*Proof:* The probabilities of detection $(P_D^d[j])$ and false alarm $(P_{FA}^d[j])$ at the destination nanomachine in the $j$th time slot can be derived using the decision rule (11) as

$$P_D^d[j] = \Pr(T(R_{sd}[j]) > \gamma'_{sd}[j]|\mathcal{H}_1)$$
$$= \Pr(R_{sd}[j] > \gamma'_{sd}[j]|\mathcal{H}_1), \quad (21)$$

$$P_{FA}^d[j] = \Pr(T(R_{sd}[j]) > \gamma'_{sd}[j]|\mathcal{H}_0)$$
$$= \Pr(R_{sd}[j] > \gamma'_{sd}[j]|\mathcal{H}_0), \quad (22)$$

where the number of received molecules $R_{sd}[j]$ is Gaussian distributed (6) under hypotheses $\mathcal{H}_0$ and $\mathcal{H}_1$ respectively. Subtracting their respective means followed by division by the standard deviations, i.e., $\frac{R_{sd}[j]-\mu_{sd,1}[j]}{\sigma_{sd,1}[j]}$ and $\frac{R_{sd}[j]-\mu_{sd,0}[j]}{\sigma_{sd,0}[j]}$ yields standard normal random variables under hypotheses $\mathcal{H}_1$ and $\mathcal{H}_0$ respectively. Subsequently, the expressions for $P_D^d[j]$ and $P_{FA}^d[j]$ in (19) and (20) respectively follow employing the definition of the $Q(\cdot)$ function $Q(\alpha) = \frac{1}{\sqrt{2\pi}}\int_\alpha^\infty e^{-\frac{x^2}{2}}dx$ [50]. ∎

### C. Probability of Error Analysis

The end-to-end probability of error for direct communication between the source and destination nanomachines follows as described in the result below.

*Theorem 3:* The average probability of error $(P_e)$ for slots 1 to $k$ at the destination nanomachine in the diffusion based cooperative molecular nano-network in which the molecules are in Brownian motion with drift, is given as

$$P_e = \frac{1}{k}\sum_{j=1}^{k}\left[\beta\left(1 - Q\left(\frac{\gamma'_{sd}[j] - \mu_{sd,1}[j]}{\sigma_{sd,1}[j]}\right)\right)\right.$$
$$\left. + (1-\beta)Q\left(\frac{\gamma'_{sd}[j] - \mu_{sd,0}[j]}{\sigma_{sd,0}[j]}\right)\right]. \quad (23)$$

*Proof:* The probability of error $P_e^d[j]$ in the $j$th time slot is defined as [50]

$$P_e^d[j] = \Pr(\text{decide } \mathcal{H}_0, \mathcal{H}_1 \text{ true}) + \Pr(\text{decide } \mathcal{H}_1, \mathcal{H}_0 \text{ true})$$
$$= (1 - P_D^d[j])\Pr(\mathcal{H}_1) + P_{FA}^d[j]\Pr(\mathcal{H}_0), \quad (24)$$

where the prior probabilities of the hypotheses $\Pr(\mathcal{H}_1)$ and $\Pr(\mathcal{H}_0)$ are $\beta$ and $1-\beta$ respectively. The quantities $P_D^d[j]$ and $P_{FA}^d[j]$ denote the probabilities of detection and false alarm at the destination nanomachine in the $j$th time slot and are given in (19) and (20) respectively. Substituting these expressions, the average probability of error for slots 1 to $k$ is obtained as stated in (23). ∎

### D. Capacity Analysis

This section presents the capacity analysis for the above system considering the impact of ISI, counting errors, and MSI. Let the discrete random variables $X[j]$ and $Y[j]$ represent the transmitted and received symbol, respectively, in the $j$th slot. The mutual information $I(X[j], Y[j])$ is

$$I(X[j], Y[j])$$
$$= \sum_{x[j]\in\{0,1\}}\sum_{y[j]\in\{0,1\}} \Pr(x[j],y[j])\log_2\frac{\Pr(y[j]|x[j])}{\Pr(y[j])}, (25)$$

where the joint and marginal probabilities $\Pr(x[j], y[j])$ and $\Pr(y[j])$ respectively can be written in terms of the conditional probability $\Pr(y[j]|x[j])$ and marginal probability $\Pr(x[j])$ as $\Pr(x[j], y[j]) = \Pr(y[j]|x[j])\Pr(x[j])$ and $\Pr(y[j]) = \sum_{x[j]\in\{0,1\}}\Pr(y[j]|x[j])\Pr(x[j])$ respectively. Substituting the above expressions for $\Pr(x[j], y[j])$ and $\Pr(y[j])$ in terms of the conditional probability $\Pr(y[j]|x[j])$ and marginal probability $\Pr(x[j])$ in (25), the mutual information $I(X[j], Y[j])$ can be obtained as (26), where the conditional probabilities $\Pr(y[j]\in\{0,1\}|x[j]\in\{0,1\})$ can be written in terms of $P_D^d[j]$ and $P_{FA}^d[j]$ as

$$\Pr(y[j]=0|x[j]=0) = 1 - P_{FA}^d[j]$$
$$= 1 - Q\left(\frac{\gamma'_{sd}[j] - \mu_{sd,0}[j]}{\sigma_{sd,0}[j]}\right),$$
$$\Pr(y[j]=1|x[j]=0) = P_{FA}^d[j]$$
$$= Q\left(\frac{\gamma'_{sd}[j] - \mu_{sd,0}[j]}{\sigma_{sd,0}[j]}\right),$$
$$\Pr(y[j]=0|x[j]=1) = 1 - P_D^d[j]$$
$$= 1 - Q\left(\frac{\gamma'_{sd}[j] - \mu_{sd,1}[j]}{\sigma_{sd,1}[j]}\right),$$
$$\Pr(y[j]=1|x[j]=1) = P_D^d[j]$$
$$= Q\left(\frac{\gamma'_{sd}[j] - \mu_{sd,1}[j]}{\sigma_{sd,1}[j]}\right).$$

The capacity $C[k]$ of the direct source-destination channel, as $k$ approaches $\infty$, can be obtained by maximizing the average mutual information $I(X[j], Y[j])$ for slots 1 to $k$ as [8]

$$C[k] = \max_\beta \frac{1}{k}\sum_{j=1}^{k} I(X[j], Y[j]) \text{ bits/slot.} \quad (27)$$

## III. DUAL-HOP COMMUNICATION BETWEEN THE SOURCE AND DESTINATION NANOMACHINES

### A. System Model

Consider now a dual-hop diffusive molecular communication system inside a blood vessel wherein the intermediate

$$I(X[j], Y[j]) = (1-\beta)\left[\Pr(y[j]=0|x[j]=0)\log_2\frac{\Pr(y[j]=0|x[j]=0)}{(1-\beta)\Pr(y[j]=0|x[j]=0)+\beta\Pr(y[j]=0|x[j]=1)}\right.$$
$$\left.+\Pr(y[j]=1|x[j]=0)\log_2\frac{\Pr(y[j]=1|x[j]=0)}{(1-\beta)\Pr(y[j]=1|x[j]=0)+\beta\Pr(y[j]=1|x[j]=1)}\right]$$
$$+\beta\left[\Pr(y[j]=0|x[j]=1)\log_2\frac{\Pr(y[j]=0|x[j]=1)}{(1-\beta)\Pr(y[j]=0|x[j]=0)+\beta\Pr(y[j]=0|x[j]=1)}\right.$$
$$\left.+\Pr(y[j]=1|x[j]=1)\log_2\frac{\Pr(y[j]=1|x[j]=1)}{(1-\beta)\Pr(y[j]=1|x[j]=0)+\beta\Pr(y[j]=1|x[j]=1)}\right], \quad (26)$$

nanomachine cooperates with the source nanomachine to relay its information to the destination nanomachine. This system model is similar to the one considered in the recent work [41] where the source, cooperative, and destination nanomachines lie on a straight line and the direction of molecular drift is from the source to cooperative and cooperative to destination nanomachines. The cooperative nanomachine is assumed to be located at a distance of $d_{sr}$ from the source nanomachine whereas the destination nanomachine is located at a distance of $d_{rd}$ from the cooperative nanomachine. Similar to the previous direct communication scenario, the channel is divided into time slots of duration $\tau$ as shown in Figs. 2 and 3, where the $j$th and $(j+1)$th slots are defined as the time intervals $[(j-1)\tau, j\tau]$ and $[j\tau, (j+1)\tau]$ respectively.

In this system, the end-to-end communication between the source and the destination nanomachines occurs in two time-slots. At the beginning of the $j$th time slot, the source nanomachine emits either $\mathcal{Q}_0[j]$ number of type 0 molecules in the propagation medium for information symbol 1 generated with a prior probability $\beta$ or remains silent for information symbol 0. The intermediate nanomachine decodes the symbol received from the source nanomachine followed by retransmission of $\mathcal{Q}_1[j+1]$ number of type 1 molecules or remaining silent in the subsequent $(j+1)$th time slot to indicate decoded symbols 1, 0 respectively. As described in [4], nanomachines such as eukaryotic cells can be genetically modified to emit different type of molecules. Further, since the source and cooperative nanomachines transmit different types of molecules, the cooperative nanomachine can operate in full duplex mode.

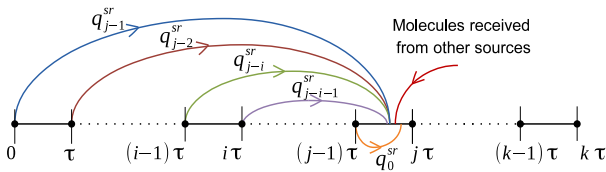

Fig. 2. Diffusion based molecular communication between the source and cooperative nanomachines over time-slotted channel, where $[(j-1)\tau, j\tau]$ is the current slot.

Let $q_{j-i}^{xy}$, $xy \in \{sr, rd\}$ denote the probability that a molecule transmitted by node $x$ in slot $i \in \{1, 2, \cdots, j\}$ arrives at node $y$ during time slot $j$. These probabilities can be obtained using (1) and setting the source-cooperative and cooperative-destination link distances to $d_{sr}$ and $d_{rd}$ respectively. The number of type 0 molecules received at the cooperative nanomachine during the time slot $[(j-1)\tau, j\tau]$ can be ex-

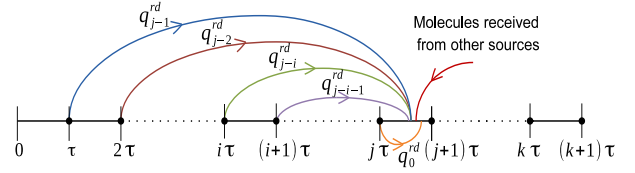

Fig. 3. Diffusion based molecular communication between the cooperative and destination nanomachines over time-slotted channel, where $[j\tau, (j+1)\tau]$ is the current slot.

pressed as
$$R_{sr}[j] = S_{sr}[j] + \mathcal{I}_{sr}[j] + N_{sr}[j] + C_{sr}[j], \quad (28)$$

where $S_{sr}[j] \sim \mathcal{N}(\mathcal{Q}_0[j]x[j]q_0^{sr}, \mathcal{Q}_0[j]x[j]q_0^{sr}(1-q_0^{sr}))$ is the number of type 0 molecules received at the cooperative nanomachine from the current slot $[(j-1)\tau, j\tau]$, with $x[j] \in \{0,1\}$ denoting the symbol transmitted by the source nanomachine in the $j$th time slot. The quantity $N_{sr}[j] \sim \mathcal{N}(\mu_o, \sigma_o^2)$ is the MSI and $C_{sr}[j] \sim \mathcal{N}(0, \mathbb{E}\{R_{sr}[j]\})$ denotes the counting errors at the cooperative nanomachine. The quantity $\mathcal{I}_{sr}[j]$ that represents the ISI at the cooperative nanomachine in slot $j$ is
$$\mathcal{I}_{sr}[j] = I_{sr}[1] + I_{sr}[2] + \cdots + I_{sr}[j-1], \quad (29)$$
where $I_{sr}[i] \sim \mathcal{N}(\mathcal{Q}_0[j-i]x[j-i]q_i^{sr}, \mathcal{Q}_0[j-i]x[j-i]q_i^{sr}(1-q_i^{sr}))$, $1 \leq i \leq j-1$ denotes the number of stray molecules received from the $(j-i)$th slot.

The cooperative nanomachine decodes the symbol using the number of received molecules $R_{sr}[j]$ followed by the retransmission of the decoded symbol $\widehat{x}[j]$. The number of type 1 molecules received at the destination nanomachine during time slot $[j\tau, (j+1)\tau]$ can be expressed as
$$R_{rd}[j+1] = S_{rd}[j+1] + \mathcal{I}_{rd}[j+1] + N_{rd}[j+1]$$
$$+ C_{rd}[j+1], \quad (30)$$
where $S_{rd}[j+1] \sim \mathcal{N}(\mathcal{Q}_1[j+1]\widehat{x}[j]q_0^{rd}, \mathcal{Q}_1[j+1]\widehat{x}[j]q_0^{rd}(1-q_0^{rd}))$ is the number of type 1 molecules received at the destination nanomachine during the current slot $[j\tau, (j+1)\tau]$. The quantities $N_{rd}[j+1] \sim \mathcal{N}(\mu_o, \sigma_o^2)$ and $C_{rd}[j+1] \sim \mathcal{N}(0, \mathbb{E}\{R_{rd}[j+1]\})$ denote the MSI and counting errors, respectively, at the destination nanomachine. Similarly, the ISI $\mathcal{I}_{rd}[j+1]$ at the destination nanomachine is
$$\mathcal{I}_{rd}[j+1] = I_{rd}[2] + I_{rd}[3] + \cdots + I_{rd}[j], \quad (31)$$
where $I_{rd}[i] \sim \mathcal{N}(\mathcal{Q}_1[j-i+2]\widehat{x}[j-i+1]q_{i-1}^{rd}, \mathcal{Q}_1[j-i+2]\widehat{x}[j-i+1]q_{i-1}^{rd}(1-q_{i-1}^{rd}))$, $2 \leq i \leq j$ denotes the number of stray molecules received from the previous $(j-i+2)$th slot.

## B. Detection Performance Analysis

On lines similar to the proof of Theorem 1, one can obtain the optimal decision rule at the cooperative nanomachine for the $j$th time slot as

$$T(R_{sr}[j]) = R_{sr}[j] \underset{\mathcal{H}_0}{\overset{\mathcal{H}_1}{\gtrless}} \gamma'_{sr}[j], \quad (32)$$

where the optimal decision threshold $\gamma'_{sr}[j]$ is given in (33). The mean $\mu_{sr,0}[j]$ and variance $\sigma^2_{sr,0}[j]$ under hypothesis $\mathcal{H}_0$ are derived using (28) as

$$\mu_{sr,0}[j] = \beta \sum_{i=1}^{j-1} \mathcal{Q}_0[j-i] q_i^{sr} + \mu_o, \quad (34)$$

$$\sigma^2_{sr,0}[j] = \sum_{i=1}^{j-1} [\beta \mathcal{Q}_0[j-i] q_i^{sr}(1-q_i^{sr}) + \beta(1-\beta) \\ \times (\mathcal{Q}_0[j-i] q_i^{sr})^2] + \sigma^2_o + \mu_{sr,0}[j], \quad (35)$$

while the mean $\mu_{sr,1}[j]$ and variance $\sigma^2_{sr,1}[j]$ under hypothesis $\mathcal{H}_1$ are

$$\mu_{sr,1}[j] = \mathcal{Q}_0[j] q_0^{sr} + \beta \sum_{i=1}^{j-1} \mathcal{Q}_0[j-i] q_i^{sr} + \mu_o, \quad (36)$$

$$\sigma^2_{sr,1}[j] = \mathcal{Q}_0[j] q_0^{sr}(1-q_0^{sr}) + \sum_{i=1}^{j-1} [\beta \mathcal{Q}_0[j-i] q_i^{sr}(1-q_i^{sr}) \\ + \beta(1-\beta)(\mathcal{Q}_0[j-i] q_i^{sr})^2] + \sigma^2_o + \mu_{sr,1}[j]. \quad (37)$$

Using the test in (32), the probabilities of detection $(P_D^{(1)}[j])$ and false alarm $(P_{FA}^{(1)}[j])$ at the cooperative nanomachine for the $j$th time slot can be derived as,

$$P_D^{(1)}[j] = Q\left(\frac{\gamma'_{sr}[j] - \mu_{sr,1}[j]}{\sigma_{sr,1}[j]}\right), \quad (38)$$

$$P_{FA}^{(1)}[j] = Q\left(\frac{\gamma'_{sr}[j] - \mu_{sr,0}[j]}{\sigma_{sr,0}[j]}\right), \quad (39)$$

where $\gamma'_{sr}[j]$ is defined in (33). Further, similar to the direct communication scenario detailed in Section II-A, the symbol detection problem at the destination nanomachine for transmission by the cooperative relay in the $(j+1)$th time slot can be formulated as the binary hypothesis testing problem

$$\begin{aligned}\mathcal{H}_0 : R_{rd}[j+1] &= I_{rd}[2] + I_{rd}[3] + \cdots + I_{rd}[j] \\ &\quad + N_{rd}[j+1] + C_{rd}[j+1] \\ \mathcal{H}_1 : R_{rd}[j+1] &= S_{rd}[j+1] + I_{rd}[2] + I_{rd}[3] + \cdots + I_{rd}[j] \\ &\quad + N_{rd}[j+1] + C_{rd}[j+1].\end{aligned} \quad (40)$$

$\mathcal{H}_0$ and $\mathcal{H}_1$ denote the null and alternative hypotheses corresponding to the transmission of decoded symbols $\widehat{x}[j]=0$ and $\widehat{x}[j]=1$ respectively by the cooperative nanomachine during the $(j+1)$th time slot. The quantities $R_{rd}[j+1]$ corresponding to the individual hypotheses in (40) are distributed as

$$\begin{aligned}\mathcal{H}_0 : R_{rd}[j+1] &\sim \mathcal{N}(\mu_{rd,0}[j+1], \sigma^2_{rd,0}[j+1]) \\ \mathcal{H}_1 : R_{rd}[j+1] &\sim \mathcal{N}(\mu_{rd,1}[j+1], \sigma^2_{rd,1}[j+1]),\end{aligned} \quad (41)$$

where the mean $\mu_{rd,0}[j+1]$ and variance $\sigma^2_{rd,0}[j+1]$ under hypothesis $\mathcal{H}_0$ are derived as,

$$\mu_{rd,0}[j+1] = \mu_I[2] + \mu_I[3] + \cdots + \mu_I[j] + \mu_o$$

$$= \beta \sum_{i=2}^{j} \mathcal{Q}_1[j-i+2] q_{i-1}^{rd} + \mu_o, \quad (42)$$

$$\sigma^2_{rd,0}[j+1] = \sigma^2_I[2] + \sigma^2_I[3] + \cdots + \sigma^2_I[j] + \sigma^2_o + \sigma^2_c[j+1]$$

$$= \sum_{i=2}^{j} [\beta \mathcal{Q}_1[j-i+2] q_{i-1}^{rd}(1-q_{i-1}^{rd}) + \beta(1-\beta) \\ \times (\mathcal{Q}_1[j-i+2] q_{i-1}^{rd})^2] + \sigma^2_o + \mu_{rd,0}[j+1], \quad (43)$$

while the mean $\mu_{rd,1}[j+1]$ and variance $\sigma^2_{rd,1}[j+1]$ under hypothesis $\mathcal{H}_1$ are derived as,

$$\mu_{rd,1}[j+1] = \mu_{rd}[j+1] + \mu_I[2] + \mu_I[3] + \cdots + \mu_I[j] + \mu_o$$

$$= \mathcal{Q}_1[j+1] q_0^{rd} + \beta \sum_{i=2}^{j} \mathcal{Q}_1[j-i+2] q_{i-1}^{rd} + \mu_o, \quad (44)$$

$$\sigma^2_{rd,1}[j+1] = \sigma^2_{rd}[j+1] + \sigma^2_I[2] + \sigma^2_I[3] + \cdots + \sigma^2_I[j] \\ + \sigma^2_o + \sigma^2_c[j+1]$$

$$= \mathcal{Q}_1[j+1] q_0^{rd}(1-q_0^{rd}) + \sum_{i=2}^{j} [\beta \mathcal{Q}_1[j-i+2] q_{i-1}^{rd} \\ \times (1-q_{i-1}^{rd}) + \beta(1-\beta)(\mathcal{Q}_1[j-i+2] q_{i-1}^{rd})^2] \\ + \sigma^2_o + \mu_{rd,1}[j+1]. \quad (45)$$

The optimal decision rule at the destination nanomachine for the dual-hop molecular communication network is obtained next.

*Theorem 4:* The optimal detector at the destination nanomachine corresponding to transmission by the cooperative nanomachine in the $(j+1)$th time slot is obtained as,

$$T(R_{rd}[j+1]) = R_{rd}[j+1] \underset{\mathcal{H}_0}{\overset{\mathcal{H}_1}{\gtrless}} \gamma'_{rd}[j+1], \quad (46)$$

with the optimal decision threshold $\gamma'_{rd}[j+1]$ given as,

$$\gamma'_{rd}[j+1] = \sqrt{\gamma_{rd}[j+1]} - \alpha_{rd}[j+1], \quad (47)$$

where the quantities $\alpha_{rd}[j+1]$ and $\gamma_{rd}[j+1]$ are defined as,

$$\alpha_{rd}[j+1] \\ = \frac{\mu_{rd,1}[j+1]\sigma^2_{rd,0}[j+1] - \mu_{rd,0}[j+1]\sigma^2_{rd,1}[j+1]}{\sigma^2_{rd,1}[j+1] - \sigma^2_{rd,0}[j+1]}, \quad (48)$$

$$\gamma_{rd}[j+1] \\ = \ln\left[\sqrt{\frac{\sigma^2_{rd,1}[j+1]}{\sigma^2_{rd,0}[j+1]}} \left\{\frac{(1-\beta)(1-P_{FA}^{(1)}[j]) - \beta(1-P_D^{(1)}[j])}{\beta P_D^{(1)}[j] - (1-\beta)P_{FA}^{(1)}[j]}\right\}\right] \\ \times \frac{2\sigma^2_{rd,1}[j+1]\sigma^2_{rd,0}[j+1]}{\sigma^2_{rd,1}[j+1] - \sigma^2_{rd,0}[j+1]} + (\alpha_{rd}[j+1])^2 \\ + \frac{\mu^2_{rd,1}[j+1]\sigma^2_{rd,0}[j+1] - \mu^2_{rd,0}[j+1]\sigma^2_{rd,1}[j+1]}{\sigma^2_{rd,1}[j+1] - \sigma^2_{rd,0}[j+1]}. \quad (49)$$

$$\gamma'_{sr}[j] = \sqrt{\frac{2\sigma^2_{sr,1}[j]\sigma^2_{sr,0}[j]}{\sigma^2_{sr,1}[j]-\sigma^2_{sr,0}[j]} \ln\left[\frac{(1-\beta)}{\beta}\sqrt{\frac{\sigma^2_{sr,1}[j]}{\sigma^2_{sr,0}[j]}}\right] + \left(\frac{\mu_{sr,1}[j]\sigma^2_{sr,0}[j]-\mu_{sr,0}[j]\sigma^2_{sr,1}[j]}{\sigma^2_{sr,1}[j]-\sigma^2_{sr,0}[j]}\right)^2 + \frac{\mu^2_{sr,1}[j]\sigma^2_{sr,0}[j]-\mu^2_{sr,0}[j]\sigma^2_{sr,1}[j]}{\sigma^2_{sr,1}[j]-\sigma^2_{sr,0}[j]}}$$
$$-\frac{\mu_{sr,1}[j]\sigma^2_{sr,0}[j]-\mu_{sr,0}[j]\sigma^2_{sr,1}[j]}{\sigma^2_{sr,1}[j]-\sigma^2_{sr,0}[j]}, \tag{33}$$

The expressions for the probabilities of detection $P_D^{(1)}[j]$ and false alarm $P_{FA}^{(1)}[j]$ for cooperative nanomachine are as obtained in (38) and (39) respectively.

*Proof:* The optimal LRT $\mathcal{L}(R_{rd}[j+1])$ at the destination nanomachine is

$$\mathcal{L}(R_{rd}[j+1]) = \frac{p(R_{rd}[j+1]|\mathcal{H}_1)}{p(R_{rd}[j+1]|\mathcal{H}_0)} \underset{\mathcal{H}_0}{\overset{\mathcal{H}_1}{\gtrless}} \frac{1-\beta}{\beta}. \tag{50}$$

The likelihood ratio $\mathcal{L}(R_{rd}[j+1])$ is evaluated in (51), where the PDFs $p(R_{rd}[j+1]|\widehat{x}[j]=1)$ and $p(R_{rd}[j+1]|\widehat{x}[j]=0)$ are

$$p(R_{rd}[j+1]|\widehat{x}[j]=1)$$
$$= \frac{1}{\sqrt{2\pi\sigma^2_{rd,1}[j+1]}} \exp\left(-\frac{(R_{rd}[j+1]-\mu_{rd,1}[j+1])^2}{2\sigma^2_{rd,1}[j+1]}\right), \tag{52}$$

$$p(R_{rd}[j+1]|\widehat{x}[j]=0)$$
$$= \frac{1}{\sqrt{2\pi\sigma^2_{rd,0}[j+1]}} \exp\left(-\frac{(R_{rd}[j+1]-\mu_{rd,0}[j+1])^2}{2\sigma^2_{rd,0}[j+1]}\right). \tag{53}$$

Substituting the above expressions followed by simplification as shown in Appendix A, the resulting test can be expressed as given in (54). The expression for the test statistic $f(R_{rd}[j+1])$ in (54) can be further simplified as shown in (55). Substituting expression for $f(R_{rd}[j+1])$ from (55) in (54) and merging the terms independent of the number of received molecules $R_{rd}[j+1]$ with the threshold in (55), the test reduces to

$$(R_{rd}[j+1] + \alpha_{rd}[j+1])^2 \underset{\mathcal{H}_0}{\overset{\mathcal{H}_1}{\gtrless}} \gamma_{rd,j+1}, \tag{56}$$

where $\alpha_{rd}[j+1]$ and $\gamma_{rd}[j+1]$ are defined in (48) and (49) respectively. Taking the square root of both sides, where $\gamma_{rd}[j+1] \geq 0$, yields the optimal test at the destination nanomachine as given in (46). It can be noted that the decision rule obtained above incorporates also the detection performance of the intermediate cooperative nanomachine. ∎ The result below determines the resulting detection performance at the destination nanomachine in the dual-hop network.

*Theorem 5:* The average probabilities of detection $P_D$ and false alarm $P_{FA}$ at the destination nanomachine corresponding to transmission by the source nanomachine in slots 1 to $k$ for the dual-hop diffusive molecular communication system are given as

$$P_D = \frac{1}{k}\sum_{j=1}^{k} P_D^d[j+1], \tag{57}$$

$$P_{FA} = \frac{1}{k}\sum_{j=1}^{k} P_{FA}^d[j+1], \tag{58}$$

where the probabilities of detection $P_D^d[j+1]$ and false alarm $P_{FA}^d[j+1]$ at the destination in the $(j+1)$th slot are given as,

$$P_D^d[j+1] = Q\left(\frac{\gamma'_{rd}[j+1]-\mu_{rd,1}[j+1]}{\sigma_{rd,1}[j+1]}\right) P_D^{(1)}[j]$$
$$+ Q\left(\frac{\gamma'_{rd}[j+1]-\mu_{rd,0}[j+1]}{\sigma_{rd,0}[j+1]}\right)(1-P_D^{(1)}[j]), \tag{59}$$

$$P_{FA}^d[j+1] = Q\left(\frac{\gamma'_{rd}[j+1]-\mu_{rd,1}[j+1]}{\sigma_{rd,1}[j+1]}\right) P_{FA}^{(1)}[j]$$
$$+ Q\left(\frac{\gamma'_{rd}[j+1]-\mu_{rd,0}[j+1]}{\sigma_{rd,0}[j+1]}\right)(1-P_{FA}^{(1)}[j]). \tag{60}$$

*Proof:* The probability of detection $P_D^d[j+1]$ at the destination nanomachine can be derived using the decision rule in (46) as

$$P_D^d[j+1]$$
$$= \Pr(T(R_{rd}[j+1]) > \gamma'_{rd}[j+1]|\mathcal{H}_1)$$
$$= \Pr(R_{rd}[j+1] > \gamma'_{rd}[j+1]|\widehat{x}[j]=1)\Pr(\widehat{x}[j]=1|\mathcal{H}_1)$$
$$+ \Pr(R_{rd}[j+1] > \gamma'_{rd}[j+1]|\widehat{x}[j]=0)\Pr(\widehat{x}[j]=0|\mathcal{H}_1)$$
$$= \Pr(R_{rd}[j+1] > \gamma'_{rd}[j+1]|\widehat{x}[j]=1)P_D^{(1)}[j]$$
$$+ \Pr(R_{rd}[j+1] > \gamma'_{rd}[j+1]|\widehat{x}[j]=0)(1-P_D^{(1)}[j]), \tag{61}$$

Similarly, the probability of false alarm $P_{FA}^d[j+1]$ can be derived as,

$$P_{FA}^d[j+1]$$
$$= \Pr(T(R_{rd}[j+1]) > \gamma'_{sd}[j+1]|\mathcal{H}_0)$$
$$= \Pr(R_{rd}[j+1] > \gamma'_{rd}[j+1]|\widehat{x}[j]=1)\Pr(\widehat{x}[j]=1|\mathcal{H}_0)$$
$$+ \Pr(R_{rd}[j+1] > \gamma'_{rd}[j+1]|\widehat{x}[j]=0)\Pr(\widehat{x}[j]=0|\mathcal{H}_0)$$
$$= \Pr(R_{rd}[j+1] > \gamma'_{rd}[j+1]|\widehat{x}[j]=1)P_{FA}^{(1)}[j]$$
$$+ \Pr(R_{rd}[j+1] > \gamma'_{rd}[j+1]|\widehat{x}[j]=0)(1-P_{FA}^{(1)}[j]). \tag{62}$$

Further, substituting $\Pr(R_{rd}[j+1] > \gamma'_{rd}[j+1]|\widehat{x}[j]=1) = Q\left(\frac{\gamma'_{rd}[j+1]-\mu_{rd,1}[j+1]}{\sigma_{rd,1}[j+1]}\right)$ and $\Pr(R_{rd}[j+1] > \gamma'_{rd}[j+1]|\widehat{x}[j]=0) = Q\left(\frac{\gamma'_{rd}[j+1]-\mu_{rd,0}[j+1]}{\sigma_{rd,0}[j+1]}\right)$ in (61) and (62), one can obtain the final expressions for $P_D^d[j+1]$ and $P_{FA}^d[j+1]$ given in (59) and (60) respectively. ∎

### C. Probability of Error Analysis

The end-to-end probability of error for dual-hop communication between the source and destination nanomachines is given by the result below.

*Theorem 6:* The average probability of error $(P_e)$ at the destination nanomachine corresponding to the transmission

$$\mathcal{L}(R_{rd}[j+1]) = \frac{p(R_{rd}[j+1]|\widehat{x}[j]=1)\Pr(\widehat{x}[j]=1|\mathcal{H}_1) + p(R_{rd}[j+1]|\widehat{x}[j]=0)\Pr(\widehat{x}[j]=0|\mathcal{H}_1)}{p(R_{rd}[j+1]|\widehat{x}[j]=1)\Pr(\widehat{x}[j]=1|\mathcal{H}_0) + p(R_{rd}[j+1]|\widehat{x}[j]=0)\Pr(\widehat{x}[j]=0|\mathcal{H}_0)}$$
$$= \frac{p(R_{rd}[j+1]|\widehat{x}[j]=1)P_D^{(1)}[j] + p(R_{rd}[j+1]|\widehat{x}[j]=0)(1-P_D^{(1)}[j])}{p(R_{rd}[j+1]|\widehat{x}[j]=1)P_{FA}^{(1)}[j] + p(R_{rd}[j+1]|\widehat{x}[j]=0)(1-P_{FA}^{(1)}[j])}, \quad (51)$$

$$\frac{1}{2\sigma_{rd,0}^2[j+1]\sigma_{rd,1}^2[j+1]} \overbrace{(R_{rd}[j+1]-\mu_{rd,0}[j+1])^2 \sigma_{rd,1}^2[j+1] - (R_{rd}[j+1]-\mu_{rd,1}[j+1])^2 \sigma_{rd,0}^2[j+1]}^{\triangleq f(R_{rd}[j+1])}$$
$$\underset{\mathcal{H}_0}{\overset{\mathcal{H}_1}{\gtrless}} \ln\left[\sqrt{\frac{\sigma_{rd,1}^2[j+1]}{\sigma_{rd,0}^2[j+1]}}\left\{\frac{(1-\beta)(1-P_{FA}^{(1)}[j]) - \beta(1-P_D^{(1)}[j])}{\beta P_D^{(1)}[j] - (1-\beta)P_{FA}^{(1)}[j]}\right\}\right]. \quad (54)$$

$$f(R_{rd}[j+1]) = R_{rd}^2[j+1](\sigma_{rd,1}^2[j+1] - \sigma_{rd,0}^2[j+1]) + 2R_{rd}[j+1](\mu_{rd,1}[j+1]\sigma_{rd,0}^2[j+1] - \mu_{rd,0}[j+1]\sigma_{rd,1}^2[j+1])$$
$$+ (\mu_{rd,0}^2[j+1]\sigma_{rd,1}^2[j+1] - \mu_{rd,1}^2[j+1]\sigma_{rd,0}^2[j+1])$$
$$= (\sigma_{rd,1}^2[j+1] - \sigma_{rd,0}^2[j+1])\left[R_{rd}[j+1] + \frac{\mu_{rd,1}[j+1]\sigma_{rd,0}^2[j+1] - \mu_{rd,0}[j+1]\sigma_{rd,1}^2[j+1]}{\sigma_{rd,1}^2[j+1] - \sigma_{rd,0}^2[j+1]}\right]^2$$
$$- \frac{(\mu_{rd,1}[j+1]\sigma_{rd,0}^2[j+1] - \mu_{rd,0}[j+1]\sigma_{rd,1}^2[j+1])^2}{\sigma_{rd,1}^2[j+1] - \sigma_{rd,0}^2[j+1]} + (\mu_{rd,1}[j+1]\sigma_{rd,0}^2[j+1] - \mu_{rd,0}[j+1]\sigma_{rd,1}^2[j+1]). \quad (55)$$

---

by the source nanomachine in slots 1 to $k$ for the dual-hop diffusive molecular communication system is given as

$$P_e = \frac{1}{k}\sum_{j=1}^{k}\left[\beta(1 - P_D^d[j+1]) + (1-\beta)P_{FA}^d[j+1]\right], \quad (63)$$

where the expressions for $P_D^d[j+1]$ and $P_{FA}^d[j+1]$ are given in (59) and (60) respectively.

*Proof:* This follows on lines similar to proof of Theorem 3. ∎

### D. Capacity Analysis

The mutual information $I(X[j], Y[j+1])$ can be expressed as

$$I(X[j], Y[j+1])$$
$$= \sum_{x[j]\in\{0,1\}}\sum_{y[j+1]\in\{0,1\}} \Pr(y[j+1]|x[j])\Pr(x[j])$$
$$\times \log_2 \frac{\Pr(y[j+1]|x[j])}{\sum_{x[j]\in\{0,1\}}\Pr(y[j+1]|x[j])\Pr(x[j])}, \quad (64)$$

where $\Pr(x[j]=1)=\beta$, $\Pr(x[j]=0)=1-\beta$, and $\Pr(y[j+1] \in \{0,1\}|x[j] \in \{0,1\})$ can be determined in terms of $(P_D^d[j+1])$ and $(P_{FA}^d[j+1])$ as

$$\Pr(y[j+1]=0|x[j]=0) = 1 - P_{FA}^d[j+1],$$
$$\Pr(y[j+1]=1|x[j]=0) = P_{FA}^d[j+1],$$
$$\Pr(y[j+1]=0|x[j]=1) = 1 - P_D^d[j+1],$$
$$\Pr(y[j+1]=1|x[j]=1) = P_D^d[j+1].$$

The capacity $C[k]$ of the dual-hop channel, as $k$ approaches $\infty$, is now obtained by maximizing the mutual information $I(X[j], Y[j+1])$

$$C[k] = \max_{\beta} \frac{1}{(k+1)}\sum_{j=1}^{k} I(X[j], Y[j+1]) \text{ bits/slot}. \quad (65)$$

The factor $\frac{1}{k+1}$ arises in the above expression in contrast to $\frac{1}{k}$ in (27) due to the fact that $k+1$ time-slots are required to transmit $k$ bits from the source to destination in the dual-hop scheme.

## IV. MULTI-HOP COMMUNICATION BETWEEN THE SOURCE AND DESTINATION NANOMACHINES

### A. System Model

Consider now a general multi-hop molecular communication system that consists of a source, destination and $N(\geq 2)$ intermediate cooperative nanomachines as shown in Fig. 4. In this system, the communication between the source and destination nanomachines occurs in a hop-by-hop fashion with $N+1$ different types of molecules. Therefore, the end-to-end communication requires a total of $N+1$ time-slots. At the beginning of the $j$th time slot, the source nanomachine either emits $\mathcal{Q}_0$ molecules of type 0 in the propagation medium for information symbol 1 generated with a prior probability $\beta$ or remains silent for information symbol 0. In the subsequent time slots, the intermediate full-duplex cooperative nanomachines decode the symbol using the number of molecules received from the previous hop nanomachine as shown in Fig. 4 and subsequently retransmit the decoded symbol to the next nanomachine. The destination nanomachine finally decodes the symbol using the number of molecules received from $N$th cooperative nanomachine ($R_N$) in the $(j+N)$th time slot.

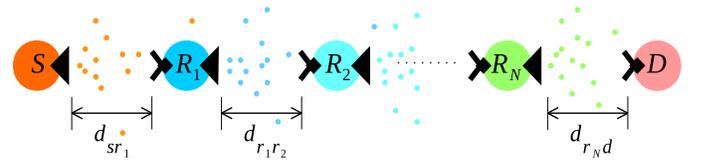

Fig. 4. Schematic diagram of multi-hop diffusion based molecular communication system with $N \geq 2$ cooperative nanomachines.

Similar to the dual-hop communication scenario, the number of molecules received at the cooperative nanomachine $R_1$ during the time slot $[(j-1)\tau, j\tau]$ can be expressed as,

$$R_{sr}[j] = S_{sr}[j] + \mathcal{I}_{sr}[j] + N_{sr}[j] + C_{sr}[j], \quad (66)$$

where the quantities $S_{sr}[j], \mathcal{I}_{sr}[j], N_{sr}[j]$, and $C_{sr}[j]$ are similar to the ones considered in (28). The expressions for the probabilities of detection $P_D^{(1)}[j]$ and false alarm $P_{FA}^{(1)}[j]$ at $R_1$ in the multi-hop system are identical to the ones obtained for the dual-hop system in (38) and (39) respectively. This is due to the fact that the system model for the received molecules at the cooperative nanomachine corresponding to the transmission by the source nanomachine in the $j$th slot is identical for both the dual-hop and multi-hop systems.

In the subsequent time slots $[(j+n-1)\tau, (j+n)\tau], n = 1, \cdots, N-1$, the number of molecules received at the cooperative nanomachine $R_{n+1}$ corresponding to the transmission of $\widehat{x}[j+n-1] \in \{0,1\}$ by the previous hop nanomachine $R_n$ can be expressed as,

$$R_{rr}[j+n] = S_{rr}[j+n] + \mathcal{I}_{rr}[j+n] + N_{rr}[j+n] + C_{rr}[j+n], \quad (67)$$

where $S_{rr}[j+n] \sim \mathcal{N}(\mathcal{Q}_n[j+n]\widehat{x}[j+n-1]q_0^{rr}, \mathcal{Q}_n[j+n]\widehat{x}[j+n-1]q_0^{rr}(1-q_0^{rr}))$ is the number of type $n$ molecules received in the current slot $[(j+n-1)\tau, (j+n)\tau]$. The quantity $N_{rr}[j+n] \sim \mathcal{N}(\mu_o, \sigma_o^2)$ is the MSI arising due to molecules received from the other sources. The term $C_{rr}[j+n] \sim \mathcal{N}(0, \mathbb{E}\{R_{rr}[j+n]\})$ denotes the counting errors at the cooperative nanomachine $R_{n+1}$. The quantity $\mathcal{I}_{rr}[j+n]$ represents the ISI at the cooperative nanomachine and is given as,

$$\mathcal{I}_{rr}[j+n] = I_{rr}[n+1] + I_{rr}[n+2] + \cdots + I_{rr}[j+n-1], \quad (68)$$

where $I_{rr}[i] \sim \mathcal{N}(\mathcal{Q}_n[j+2n-i]\widehat{x}[j+2n-i-1]q_{i-n}^{rr}, \mathcal{Q}_n[j+2n-i]\widehat{x}[j+2n-i-1]q_{i-n}^{rr}(1-q_{i-n}^{rr})), n+1 \leq i \leq j+n-1$ denotes the number of stray molecules received from the previous $(j+2n-i)$th slot.

The number of molecules received at the destination nanomachine during the time slot $[(j+N-1)\tau, (j+N)\tau]$ can be expressed as,

$$R_{rd}[j+N] = S_{rd}[j+N] + \mathcal{I}_{rd}[j+N] + N_{rd}[j+N] + C_{rd}[j+N], \quad (69)$$

where $S_{rd}[j+N] \sim \mathcal{N}(\mathcal{Q}_N[j+N]\widehat{x}[j+N-1]q_0^{rd}, \mathcal{Q}_N[j+N]\widehat{x}[j+N-1]q_0^{rd}(1-q_0^{rd}))$ is the number of type $N$ molecules received at the destination nanomachine in the current slot $[(j+N-1)\tau, (j+N)\tau]$. The quantity $N_{rd}[j+N] \sim \mathcal{N}(\mu_o, \sigma_o^2)$ is the MSI. The term $C_{rd}[j+N] \sim \mathcal{N}(0, \mathbb{E}\{R_{rd}[j+N]\})$ denotes the counting errors at the destination nanomachine. The quantity $\mathcal{I}_{rd}[j+N]$ represents the ISI at the destination nanomachine that is given as,

$$\mathcal{I}_{rd}[j+N] = I_{rd}[N+1] + I_{rd}[N+2] + \cdots + I_{rd}[j+N-1], \quad (70)$$

where $I_{rd}[i] \sim \mathcal{N}(\mathcal{Q}_N[j+2N-i]\widehat{x}[j+2N-i-1]q_{i-N}^{rd}, \mathcal{Q}_N[j+2N-i]\widehat{x}[j+2N-i-1]q_{i-N}^{rd}(1-q_{i-N}^{rd}))$, $N+1 \leq i \leq j+N-1$ denotes the number of stray molecules received from transmission during the previous $(j+2N-i)$th slot.

### B. Detection Performance Analysis

The binary hypothesis testing problem for symbol detection at the destination nanomachine corresponding to the transmission of cooperative nanomachine $R_N$ in the $(j+N)$th time slot can again be formulated as

$$\begin{aligned}\mathcal{H}_0 : & R_{rd}[j+N] = I_{rd}[N+1] + I_{rd}[N+2] + \cdots + I_{rd}[j+N-1] \\ & + N_{rd}[j+N] + C_{rd}[j+N] \\ \mathcal{H}_1 : & R_{rd}[j+N] = S_{rd}[j+N] + I_{rd}[N+1] + I_{rd}[N+2] + \cdots \\ & + I_{rd}[j+N-1] + N_{rd}[j+N] + C_{rd}[j+N],\end{aligned} \quad (71)$$

where $\mathcal{H}_0$ and $\mathcal{H}_1$ denote the null and alternative hypotheses corresponding to the transmission of the decoded symbol $\widehat{x}[j+N-1] = 0$ and $\widehat{x}[j+N-1] = 1$ respectively by the $R_N$ in the $(j+N)$th time slot. The number of molecules received at the destination nanomachine corresponding to the individual hypotheses in (71) are Gaussian distributed as

$$\begin{aligned}\mathcal{H}_0 : R_{rd}[j+N] \sim \mathcal{N}(\mu_{rd,0}[j+N], \sigma_{rd,0}^2[j+N]) \\ \mathcal{H}_1 : R_{rd}[j+N] \sim \mathcal{N}(\mu_{rd,1}[j+N], \sigma_{rd,1}^2[j+N]),\end{aligned} \quad (72)$$

where the mean $\mu_{rd,0}[j+N]$ and variance $\sigma_{rd,0}^2[j+N]$ under hypothesis $\mathcal{H}_0$ are

$$\begin{aligned}\mu_{rd,0}[j+N] &= \mu_I[N+1] + \mu_I[N+2] + \cdots + \mu_I[j+N-1] + \mu_o \\ &= \beta \sum_{i=N+1}^{j+N-1} \mathcal{Q}_N[j+2N-i]q_{i-N}^{rd} + \mu_o,\end{aligned} \quad (73)$$

$$\begin{aligned}\sigma_{rd,0}^2[j+N] &= \sigma_I^2[N+1] + \sigma_I^2[N+2] + \cdots + \sigma_I^2[j+N-1] \\ &\quad + \sigma_o^2 + \sigma_c^2[j+N] \\ &= \sum_{i=N+1}^{j+N-1} [\beta \mathcal{Q}_N[j+2N-i]q_{i-N}^{rd}(1-q_{i-N}^{rd}) + \beta \\ &\quad \times (1-\beta)(\mathcal{Q}_N[j+2N-i]q_{i-N}^{rd})^2] \\ &\quad + \sigma_o^2 + \mu_{rd,0}[j+N],\end{aligned} \quad (74)$$

and the mean $\mu_{rd,1}[j+1]$ and variance $\sigma_{rd,1}^2[j+1]$ under hypothesis $\mathcal{H}_1$ are

$$\begin{aligned}\mu_{rd,1}[j+N] &= \mu_{rd}[j+N] + \mu_I[N+1] + \mu_I[N+2] \\ &\quad + \cdots + \mu_I[j+N-1] + \mu_o \\ &= \mathcal{Q}_N[j+N]q_0^{rd} + \beta \sum_{i=N+1}^{j+N-1} \mathcal{Q}_N[j+2N-i]q_{i-N}^{rd} \\ &\quad + \mu_o,\end{aligned} \quad (75)$$

$$\begin{aligned}\sigma_{rd,1}^2[j+N] &= \sigma_{rd}^2[j+N] + \sigma_I^2[N+1] + \sigma_I^2[N+2] \\ &\quad + \cdots + \sigma_I^2[j+N-1] + \sigma_o^2 + \sigma_c^2[j+N] \\ &= \mathcal{Q}_N[j+N]q_0^{rd}(1-q_0^{rd}) \\ &\quad + \sum_{i=N+1}^{j+N-1} [\beta \mathcal{Q}_N[j+2N-i]q_{i-N}^{rd}(1-q_{i-N}^{rd}) \\ &\quad + \beta(1-\beta)(\mathcal{Q}_N[j+2N-i]q_{i-N}^{rd})^2] \\ &\quad + \sigma_o^2 + \mu_{rd,1}[j+N].\end{aligned} \quad (76)$$

For the above multi-hop scheme, the optimal test at the destination nanomachine in a multi-hop molecular communication network is given by the result below.

*Theorem 7:* The optimal detector at the destination nanomachine corresponding to the received signal in the $(j+N)$th time slot is

$$T(R_{rd}[j+N]) = R_{rd}[j+N] \underset{\mathcal{H}_0}{\overset{\mathcal{H}_1}{\gtrless}} \gamma'_{rd}[j+N], \quad (77)$$

where the optimal decision threshold $\gamma'_{rd}[j+N]$ is given as,

$$\gamma'_{rd}[j+N] = \sqrt{\gamma_{rd}[j+N]} - \alpha_{rd}[j+N]. \quad (78)$$

The quantity $\alpha_{rd}[j+N]$ is

$$\alpha_{rd}[j+N] = \frac{\mu_{rd,1}[j+N]\sigma^2_{rd,0}[j+N] - \mu_{rd,0}[j+N]\sigma^2_{rd,1}[j+N]}{\sigma^2_{rd,1}[j+N] - \sigma^2_{rd,0}[j+N]}, \quad (79)$$

with $\gamma_{rd}[j+N]$ as given in (80) and $\beta_{rd}$ is defined as

$$\beta_{rd} = \frac{(1-\beta)(1-P^{(N)}_{FA}[j+N-1]) - \beta(1-P^{(N)}_{d}[j+N-1])}{\beta P^{(N)}_{d}[j+N-1] - (1-\beta)P^{(N)}_{FA}[j+N-1]}, \quad (81)$$

The associated expressions for the probabilities of detection $P^{(N)}_{D}[j+N-1]$ and false alarm $P^{(N)}_{FA}[j+N-1]$ for the $N$th cooperative nanomachine can be obtained as given in (116) and (117) respectively.

*Proof:* Let $\xi_l(n)$ denote the state of the $n$th cooperative nanomachine $(R_n)$ with $\xi_l(n) = 1$ if $R_n$ decodes the source information symbol as 1 and $\xi_l(n) = 0$ otherwise. Let $\xi_l = [\xi_l(1), \xi_l(2), \cdots, \xi_l(N)], 0 \le l \le 2^N - 1$ denote the set of all the possible $2^N$ binary vector states. The binary vector $\xi_0 = [0, 0, \cdots, 0, 0]$ represents the state when all the cooperative nanomachines decode the source symbol as 0 whereas $\xi_{2^K-1} = [1, 1, \cdots, 1, 1]$ represents the state when all the cooperative nanomachines decode the symbol as 1. Further, let the set $\Psi_l$ defined as $\Psi_l = \{n|\xi_l(n) = 1, n = 1, 2, \cdots, N\}$ include all the cooperative nanomachines which decode the symbol as 1. Employing the above notation, the optimal LRT $\mathcal{L}(R_{rd}[j+N])$ at the destination nanomachine can be obtained as,

$$\mathcal{L}(R_{rd}[j+N]) = \frac{p(R_{rd}[j+N]|\mathcal{H}_1)}{p(R_{rd}[j+N]|\mathcal{H}_0)} \underset{\mathcal{H}_0}{\overset{\mathcal{H}_1}{\gtrless}} \frac{1-\beta}{\beta}. \quad (82)$$

The test statistic $\mathcal{L}(R_{rd}[j+N])$ above is further simplified in (83), where $p(R_{rd}[j+N]|\xi_l)$ denotes the PDF of the received molecules $R_{rd}[j+N]$ at the destination nanomachine for the scenario when the network is in state $\xi_l$ and is given as,

$$p(R_{rd}[j+N]|\xi_l) = \begin{cases} \mathcal{N}(\mu_{rd,0}[j+N], \sigma^2_{rd,0}[j+N]) & \text{if } l=0,2,\cdots,2^N-2, \\ \mathcal{N}(\mu_{rd,1}[j+N], \sigma^2_{rd,1}[j+N]) & \text{if } l=1,3,\cdots,2^N-1, \end{cases} \quad (84)$$

The quantity $\Pr(\xi_l|\mathcal{H}_i), i \in \{0,1\}$ in (83) represents the conditional probability that the network is in state $\xi_l$. Since the source and each of the cooperative nanomachines employ different types of molecules for transmission, the probability $\Pr(\xi_l|\mathcal{H}_i)$ of the system being in state $\xi_l$ under hypothesis $\mathcal{H}_i$ follows as,

$$\Pr(\xi_l|\mathcal{H}_i) = \prod_{n=1}^{N} \Pr(\xi_l(n)|\mathcal{H}_i), \quad (85)$$

where the probabilities $\Pr(\xi_l(n)|\mathcal{H}_0)$ and $\Pr(\xi_l(n)|\mathcal{H}_1)$ of the cooperative nanomachine being in state $\xi_l(n)$ under $\mathcal{H}_0$ and $\mathcal{H}_1$ can be expressed as,

$$\Pr(\xi_l(n)|\mathcal{H}_0) = \begin{cases} P^{(n)}_{FA}[j+n-1], & \text{if } n \in \Psi_l, \\ 1 - P^{(n)}_{FA}[j+n-1], & \text{if } n \in \overline{\Psi}_l, \end{cases} \quad (86)$$

$$\Pr(\xi_l(n)|\mathcal{H}_1) = \begin{cases} P^{(n)}_{D}[j+n-1], & \text{if } n \in \Psi_l, \\ 1 - P^{(n)}_{D}[j+n-1], & \text{if } n \in \overline{\Psi}_l. \end{cases} \quad (87)$$

The set $\overline{\Psi}_l$ comprises of all the cooperative nanomachines that decode the symbol as 0, and $P^{(n)}_{D}[j+n-1]$ and $P^{(n)}_{FA}[j+n-1]$ are the probabilities of detection and false alarm of the $n$th intermediate cooperative nanomachine. Employing the above expressions, the probabilities of the system being in states $\xi_l$ under $\mathcal{H}_0$ and $\mathcal{H}_1$ can be written as,

$$\Pr(\xi_l|\mathcal{H}_0) = \prod_{n \in \Psi_l} P^{(n)}_{FA}[j+n-1] \prod_{n \in \overline{\Psi}_l} \left(1 - P^{(n)}_{FA}[j+n-1]\right), \quad (88)$$

$$\Pr(\xi_l|\mathcal{H}_1) = \prod_{n \in \Psi_l} P^{(n)}_{D}[j+n-1] \prod_{n \in \overline{\Psi}_l} \left(1 - P^{(n)}_{D}[j+n-1]\right). \quad (89)$$

Further, substituting the Gaussian PDFs from (84) in (83), the test can be expressed as in (90), where $\beta_{rd}$ is defined as,

$$\beta_{rd} = \frac{\sum_{l=0,2,\cdots,2^N-2}[(1-\beta)\Pr(\xi_l|\mathcal{H}_0) - \beta\Pr(\xi_l|\mathcal{H}_1)]}{\sum_{l=1,3,\cdots,2^N-1}[\beta\Pr(\xi_l|\mathcal{H}_1) - (1-\beta)\Pr(\xi_l|\mathcal{H}_0)]},$$

$$= \frac{(1-\beta)(1-P^{(N)}_{FA}[j+N-1]) - \beta(1-P^{(N)}_{D}[j+N-1])}{\beta P^{(N)}_{D}[j+N-1] - (1-\beta)P^{(N)}_{FA}[j+N-1]}. \quad (91)$$

The detailed derivations for the test (90) and $\beta_{rd}$ above are given in Appendix B. The expression for $f(R_{rd}[j+N])$ in (90) can be further simplified as shown in (92). Substituting the resulting expression for $f(R_{rd}[j+N])$ in (90) and subsequently merging the terms independent of the received molecules $R_{rd}[j+N]$ with the threshold yields the test

$$(R_{rd}[j+N] + \alpha_{rd}[j+N])^2 \underset{\mathcal{H}_0}{\overset{\mathcal{H}_1}{\gtrless}} \gamma_{rd}[j+N], \quad (93)$$

where $\alpha_{rd}[j+N]$ and $\gamma_{rd}[j+N]$ are defined in (79) and (80) respectively. Finally, taking the square root of both sides with $\gamma_{rd}[j+N] \ge 0$, the above expression can be simplified to yield the optimal test in (77). ∎

Result below characterizes the detection performance at the destination nanomachine in the multi-hop network.

*Theorem 8:* The average probabilities of detection $P_D$ and false alarm $P_{FA}$ at the destination nanomachine corresponding to the transmission by the source nanomachine in slots 1 to $k$ in the multi-hop molecular communication system with $N \ge 2$ cooperative nanomachines are given as,

$$P_D = \frac{1}{k}\sum_{j=1}^{k} P^d_D[j+N], \quad (94)$$

$$P_{FA} = \frac{1}{k}\sum_{j=1}^{k} P^d_{FA}[j+N], \quad (95)$$

$$\gamma_{rd}[j+N] = \ln\left[\sqrt{\frac{\sigma_{rd,1}^2[j+N]}{\sigma_{rd,0}^2[j+N]}}\beta_{rd}\right] \frac{2\sigma_{rd,1}^2[j+N]\sigma_{rd,0}^2[j+N]}{\sigma_{rd,1}^2[j+N]-\sigma_{rd,0}^2[j+N]} + (\alpha_{rd}[j+N])^2$$
$$+ \frac{\mu_{rd,1}^2[j+N]\sigma_{rd,0}^2[j+N] - \mu_{rd,0}^2[j+N]\sigma_{rd,1}^2[j+N]}{\sigma_{rd,1}^2[j+N]-\sigma_{rd,0}^2[j+N]}, \quad (80)$$

$$\mathcal{L}(R_{rd}[j+N]) = \frac{\sum_{l=0}^{2^N-1} p(R_{rd}[j+N]|\xi_l)\text{Pr}(\xi_l|\mathcal{H}_1)}{\sum_{l=0}^{2^N-1} p(R_{rd}[j+N]|\xi_l)\text{Pr}(\xi_l|\mathcal{H}_0)}$$
$$= \frac{\sum_{l=0,2,\cdots,2^N-2} p(R_{rd}[j+N]|\xi_l)\text{Pr}(\xi_l|\mathcal{H}_1) + \sum_{l=1,3,\cdots,2^N-1} p(R_{rd}[j+N]|\xi_l)\text{Pr}(\xi_l|\mathcal{H}_1)}{\sum_{l=0,2,\cdots,2^N-2} p(R_{rd}[j+N]|\xi_l)\text{Pr}(\xi_l|\mathcal{H}_0) + \sum_{l=1,3,\cdots,2^N-1} p(R_{rd}[j+N]|\xi_l)\text{Pr}(\xi_l|\mathcal{H}_0)}, \quad (83)$$

$$\overbrace{\frac{(R_{rd}[j+N]-\mu_{rd,0}[j+N])^2\sigma_{rd,1}^2[j+N] - (R_{rd}[j+N]-\mu_{rd,1}[j+N])^2\sigma_{rd,0}^2[j+N]}{2\sigma_{rd,0}^2[j+N]\sigma_{rd,1}^2[j+N]}}^{\triangleq f(R_{rd}[j+N])} \underset{\mathcal{H}_0}{\overset{\mathcal{H}_1}{\gtrless}} \ln\left[\sqrt{\frac{\sigma_{rd,1}^2[j+N]}{\sigma_{rd,0}^2[j+N]}}\beta_{rd}\right]. \quad (90)$$

$$f(R_{rd}[j+N])$$
$$= R_{rd}^2[j+N](\sigma_{rd,1}^2[j+N]-\sigma_{rd,0}^2[j+N]) + 2R_{rd}[j+N](\mu_{rd,1}[j+N]\sigma_{rd,0}^2[j+N] - \mu_{rd,0}[j+N]\sigma_{rd,1}^2[j+N])$$
$$+ (\mu_{rd,0}^2[j+N]\sigma_{rd,1}^2[j+N] - \mu_{rd,1}^2[j+N]\sigma_{rd,0}^2[j+N])$$
$$= (\sigma_{rd,1}^2[j+N]-\sigma_{rd,0}^2[j+N])\left[R_{rd}[j+N] + \frac{\mu_{rd,1}[j+N]\sigma_{rd,0}^2[j+N] - \mu_{rd,0}[j+N]\sigma_{rd,1}^2[j+N]}{\sigma_{rd,1}^2[j+N]-\sigma_{rd,0}^2[j+N]}\right]^2$$
$$- \frac{(\mu_{rd,1}[j+N]\sigma_{rd,0}^2[j+N] - \mu_{rd,0}[j+N]\sigma_{rd,1}^2[j+N])^2}{\sigma_{rd,1}^2[j+N]-\sigma_{rd,0}^2[j+N]} + (\mu_{rd,1}[j+N]\sigma_{rd,0}^2[j+N] - \mu_{rd,0}[j+N]\sigma_{rd,1}^2[j+N]). \quad (92)$$

where the probabilities of detection $P_D^d[j+N]$ and false alarm $P_{FA}^d[j+N]$ at the destination in the $(j+N)$th slot are given as,

$$P_D^d[j+N]$$
$$= Q\left(\frac{\gamma_{rd}'[j+N]-\mu_{rd,0}[j+N]}{\sigma_{rd,0}[j+N]}\right)\left(1-P_D^{(N)}[j+N-1]\right)$$
$$+ Q\left(\frac{\gamma_{rd}'[j+N]-\mu_{rd,1}[j+N]}{\sigma_{rd,1}[j+N]}\right)P_D^{(N)}[j+N-1], \quad (96)$$

$$P_{FA}^d[j+N]$$
$$= Q\left(\frac{\gamma_{rd}'[j+N]-\mu_{rd,0}[j+N]}{\sigma_{rd,0}[j+N]}\right)\left(1-P_{FA}^{(N)}[j+N-1]\right)$$
$$+ Q\left(\frac{\gamma_{rd}'[j+N]-\mu_{rd,1}[j+N]}{\sigma_{rd,1}[j+N]}\right)P_{FA}^{(N)}[j+N-1], \quad (97)$$

with the threshold $\gamma_{rd}'[j+N]$ as defined in (78).

*Proof:* The probability of detection $P_D^d[j+N]$ at the destination nanomachine corresponding to the transmission by $R_N$ in the $(j+N)$th slot can be derived using the test statistic $T(R_{rd}[j+N])$ given in (77) as

$$P_D^d[j+N]$$
$$= \text{Pr}(T(R_{rd}[j+N]) > \gamma_{rd}'[j+N]|\mathcal{H}_1)$$
$$= \sum_{l=0}^{2^N-1} \text{Pr}(R_{rd}[j+N] > \gamma_{rd}'[j+N]|\xi_l)\text{Pr}(\xi_l|\mathcal{H}_1), \quad (98)$$

where $\text{Pr}(\xi_l|\mathcal{H}_1)$ is given in (89) and $\text{Pr}(R_{rd}[j+N] > \gamma_{rd}'[j+N]|\xi_l)$ can be obtained using (84) as,

$$\text{Pr}(R_{rd}[j+N] > \gamma_{rd}'[j+N]|\xi_l)$$
$$= \begin{cases} Q\left(\frac{\gamma_{rd}'[j+N]-\mu_{rd,0}[j+N]}{\sigma_{rd,0}[j+N]}\right) & \text{if } l=0,2,\cdots,2^N-2, \\ Q\left(\frac{\gamma_{rd}'[j+N]-\mu_{rd,1}[j+N]}{\sigma_{rd,1}[j+N]}\right) & \text{if } l=1,3,\cdots,2^N-1. \end{cases} \quad (99)$$

Finally, employing (89) and (99) in (98), the expression for $P_D^d[j+N]$ is obtained as

$$P_D^d[j+N]$$
$$= \sum_{l=0,2,\cdots,2^N-2} Q\left(\frac{\gamma_{rd}'[j+N]-\mu_{rd,0}[j+N]}{\sigma_{rd,0}[j+N]}\right)$$
$$\times \left[\prod_{n\in\Psi_l}P_D^{(n)}[j+n-1]\prod_{n\in\overline{\Psi}_l}\left(1-P_D^{(n)}[j+n-1]\right)\right]$$
$$+ \sum_{l=1,3,\cdots,2^N-1} Q\left(\frac{\gamma_{rd}'[j+N]-\mu_{rd,1}[j+N]}{\sigma_{rd,1}[j+N]}\right)$$
$$\times \left[\prod_{n\in\Psi_l}P_D^{(n)}[j+n-1]\prod_{n\in\overline{\Psi}_l}\left(1-P_D^{(n)}[j+n-1]\right)\right], \quad (100)$$

The above expression for $P_D^d[j+N]$ considering $2^N$ possible states for $N$ cooperative nanomachines reduces to (96).

Similarly, the probability of false alarm $P_{FA}^d[j+N]$ at the destination nanomachine in the $(j+N)$th slot can be derived

as
$$P_{FA}^d[j+N]$$
$$=\Pr(T(R_{rd}[j+N]) > \gamma'_{rd}[j+N]|\mathcal{H}_0)$$
$$= \sum_{l=0}^{2^N-1} \Pr(R_{rd}[j+N] > \gamma'_{rd}[j+N]|\xi_l)\Pr(\xi_l|\mathcal{H}_0), \quad (101)$$

where $\Pr(\xi_l|\mathcal{H}_0)$ is given in (88). Further, substituting (88) and (99) in (101), the expression for $P_{FA}^d[j+N]$ follows as

$$P_{FA}^d[j+N]$$
$$= \sum_{l=0,2,\cdots,2^N-2} Q\left(\frac{\gamma'_{rd}[j+N]-\mu_{rd,0}[j+N]}{\sigma_{rd,0}[j+N]}\right)$$
$$\times \left[\prod_{n\in\Psi_l} P_{FA}^{(n)}[j+n-1] \prod_{n\in\overline{\Psi}_l}\left(1-P_{FA}^{(n)}[j+n-1]\right)\right]$$
$$+ \sum_{l=1,3,\cdots,2^N-1} Q\left(\frac{\gamma'_{rd}[j+N]-\mu_{rd,1}[j+N]}{\sigma_{rd,1}[j+N]}\right)$$
$$\times \left[\prod_{n\in\Psi_l} P_{FA}^{(n)}[j+n-1] \prod_{n\in\overline{\Psi}_l}\left(1-P_{FA}^{(n)}[j+n-1]\right)\right]. \quad (102)$$

Similar to proof for $P_D^d[j+N]$ shown above, the expression in (102) can be simplified considering $2^N$ possible states for $N$ cooperative nanomachines as (97). ∎

Similar to the destination nanomachine, the binary hypothesis testing problem for symbol detection at the cooperative nanomachine $R_{n+1}, n = 1, 2, \cdots, N-1$, corresponding to the transmission of the cooperative nanomachine $R_n$ in $(j+n-1)$th time slot can also be formulated as

$$\mathcal{H}_0: R_{rr}[j+n] = I_{rr}[n+1] + I_{rr}[n+2] + \cdots + I_{rr}[j+n-1]$$
$$+ N_{rr}[j+n] + C_{rr}[j+n]$$
$$\mathcal{H}_1: R_{rr}[j+n] = S_{rr}[j+n] + I_{rr}[n+1] + I_{rr}[n+2] + \cdots$$
$$+ I_{rr}[j+n-1] + N_{rr}[j+n] + C_{rr}[j+n], \quad (103)$$

where $\mathcal{H}_0$ and $\mathcal{H}_1$ denote the null and alternative hypotheses corresponding to the transmission of the decoded symbol $\widehat{x}[j+n-1] = 0$ and $\widehat{x}[j+n-1] = 1$ respectively by $R_n$ in the $(j+n-1)$th time slot. The number of molecules received at $R_{n+1}$ corresponding to the individual hypotheses in (103) are Gaussian distributed as

$$\mathcal{H}_0 : R_{rr}[j+n] \sim \mathcal{N}(\mu_{rr,0}[j+n], \sigma_{rr,0}^2[j+n])$$
$$\mathcal{H}_1 : R_{rr}[j+n] \sim \mathcal{N}(\mu_{rr,1}[j+n], \sigma_{rr,1}^2[j+n]), \quad (104)$$

where the mean $\mu_{rr,0}[j+n]$ and variance $\sigma_{rr,0}^2[j+n]$ under hypothesis $\mathcal{H}_0$ are given as,

$$\mu_{rr,0}[j+n] = \beta \sum_{i=n+1}^{j+n-1} \mathcal{Q}_n[j+2n-i]q_{i-n}^{rr} + \mu_o, \quad (105)$$
$$\sigma_{rr,0}^2[j+n] = \sum_{i=n+1}^{j+n-1}[\beta\mathcal{Q}_n[j+2n-i]q_{i-n}^{rr}(1-q_{i-n}^{rr}) + \beta$$
$$\times (1-\beta)(\mathcal{Q}_n[j+2n-i]q_{i-n}^{rr})^2]$$
$$+ \sigma_o^2 + \mu_{rr,0}[j+n], \quad (106)$$

and the mean $\mu_{rr,1}[j+n]$ and variance $\sigma_{rr,1}^2[j+n]$ under hypothesis $\mathcal{H}_1$ are given as,

$$\mu_{rr,1}[j+n] = \mathcal{Q}_n[j+n]q_0^{rr} + \beta\sum_{i=n+1}^{j+n-1}\mathcal{Q}_n[j+2n-i]q_{i-n}^{rr}$$
$$+ \mu_o, \quad (107)$$
$$\sigma_{rr,1}^2[j+n] = \mathcal{Q}_n[j+n]q_0^{rr}(1-q_0^{rr})$$
$$+ \sum_{i=n+1}^{j+n-1}[\beta\mathcal{Q}_n[j+2n-i]q_{i-n}^{rr}(1-q_{i-n}^{rr})$$
$$+ \beta(1-\beta)(\mathcal{Q}_n[j+2n-i]q_{i-n}^{rr})^2]$$
$$+ \sigma_o^2 + \mu_{rr,1}[j+n]. \quad (108)$$

Hence, the optimal test for deciding 0 and 1 at the cooperative nanomachine $R_{n+1}$ considering all possible $2^n$ states of previous $n$ cooperative nanomachines i.e., $R_1, R_2, \cdots, R_n$ can be obtained as

$$T(R_{rr}[j+n]) = R_{rr}[j+n] \underset{\mathcal{H}_0}{\overset{\mathcal{H}_1}{\gtrless}} \gamma'_{rr}[j+n], \quad (109)$$

where the optimal decision threshold $\gamma'_{rr}[j+n]$ is

$$\gamma'_{rr}[j+n] = \sqrt{\gamma_{rr}[j+n]} - \alpha_{rr}[j+n], \quad (110)$$

where

$$\alpha_{rr}[j+n]$$
$$= \frac{\mu_{rr,1}[j+n]\sigma_{rr,0}^2[j+n] - \mu_{rr,0}[j+n]\sigma_{rr,1}^2[j+n]}{\sigma_{rr,1}^2[j+n] - \sigma_{rr,0}^2[j+n]}, \quad (111)$$

and $\gamma_{rr}[j+n]$ is given in (112), where $\beta_{rr}$ is defined as

$$\beta_{rr} = \frac{(1-\beta)(1-P_{FA}^{(n)}[j+n-1]) - \beta(1-P_d^{(n)}[j+n-1])}{\beta P_d^{(n)}[j+n-1] - (1-\beta)P_{FA}^{(n)}[j+n-1]}. \quad (113)$$

The quantities $P_D^{(n)}[j+n-1]$ and $P_{FA}^{(n)}[j+n-1]$ denote the probabilities of detection and false alarm for the $n$th cooperative nanomachine. Further, the probabilities of detection $P_D^{(n+1)}[j+n]$ and false alarm $P_{FA}^{(n+1)}[j+n]$ at the cooperative nanomachine $R_{n+1}$ corresponding to the transmission by $R_n$ in the $(j+n)$th slot can be derived using the test statistic $T(R_{rr}[j+n])$ in (109) as

$$P_D^{(n+1)}[j+n]$$
$$= \Pr(T(R_{rr}[j+n]) > \gamma'_{rr}[j+n]|\mathcal{H}_1)$$
$$= \sum_{l=0}^{2^n-1}\Pr(R_{rr}[j+n] > \gamma'_{rr}[j+n]|\xi_l)\Pr(\xi_l|\mathcal{H}_1), \quad (114)$$
$$P_{FA}^{(n+1)}[j+n]$$
$$= \Pr(T(R_{rr}[j+n]) > \gamma'_{rr}[j+n]|\mathcal{H}_0)$$
$$= \sum_{l=0}^{2^n-1}\Pr(R_{rr}[j+n] > \gamma'_{rr}[j+n]|\xi_l)\Pr(\xi_l|\mathcal{H}_0), \quad (115)$$

$$\gamma_{rr}[j+n] = \ln\left[\sqrt{\frac{\sigma_{rr,1}^2[j+n]}{\sigma_{rr,0}^2[j+n]}}\beta_{rr}\right]\frac{2\sigma_{rr,1}^2[j+n]\sigma_{rr,0}^2[j+n]}{\sigma_{rr,1}^2[j+n]-\sigma_{rr,0}^2[j+n]}+(\alpha_{rr}[j+n])^2$$
$$+\frac{\mu_{rr,1}^2[j+n]\sigma_{rr,0}^2[j+n]-\mu_{rr,0}^2[j+n]\sigma_{rr,1}^2[j+n]}{\sigma_{rr,1}^2[j+n]-\sigma_{rr,0}^2[j+n]}, \quad (112)$$

On lines similar to proof of Theorem 8, the above expressions $P_D^{(n+1)}[j+n]$ and $P_{FA}^{(n+1)}[j+n]$ can be simplified as

$$P_D^{(n+1)}[j+n]$$
$$=\sum_{l=0,2,\cdots,2^n-2} Q\left(\frac{\gamma'_{rr}[j+n]-\mu_{rr,0}[j+n]}{\sigma_{rr,0}[j+n]}\right)$$
$$\times \left[\prod_{n\in\Psi_l}P_D^{(n)}[j+n-1]\prod_{n\in\overline{\Psi}_l}\left(1-P_D^{(n)}[j+n-1]\right)\right]$$
$$+\sum_{l=1,3,\cdots,2^n-1} Q\left(\frac{\gamma'_{rr}[j+n]-\mu_{rr,1}[j+n]}{\sigma_{rr,1}[j+n]}\right)$$
$$\times \left[\prod_{n\in\Psi_l}P_D^{(n)}[j+n-1]\prod_{n\in\overline{\Psi}_l}\left(1-P_D^{(n)}[j+n-1]\right)\right]$$
$$=Q\left(\frac{\gamma'_{rr}[j+n]-\mu_{rr,0}[j+n]}{\sigma_{rr,0}[j+n]}\right)\left(1-P_D^{(n)}[j+n-1]\right)$$
$$+Q\left(\frac{\gamma'_{rr}[j+n]-\mu_{rr,1}[j+n]}{\sigma_{rr,1}[j+n]}\right)P_D^{(n)}[j+n-1]. \quad (116)$$

$$P_{FA}^{(n+1)}[j+n]$$
$$=\sum_{l=0,2,\cdots,2^n-2} Q\left(\frac{\gamma'_{rr}[j+n]-\mu_{rr,0}[j+n]}{\sigma_{rr,0}[j+n]}\right)$$
$$\times \left[\prod_{n\in\Psi_l}P_{FA}^{(n)}[j+n-1]\prod_{n\in\overline{\Psi}_l}\left(1-P_{FA}^{(n)}[j+n-1]\right)\right]$$
$$+\sum_{l=1,3,\cdots,2^n-1} Q\left(\frac{\gamma'_{rr}[j+n]-\mu_{rr,1}[j+n]}{\sigma_{rr,1}[j+n]}\right)$$
$$\times \left[\prod_{n\in\Psi_l}P_{FA}^{(n)}[j+n-1]\prod_{n\in\overline{\Psi}_l}\left(1-P_{FA}^{(n)}[j+n-1]\right)\right]$$
$$=Q\left(\frac{\gamma'_{rr}[j+n]-\mu_{rr,0}[j+n]}{\sigma_{rr,0}[j+n]}\right)\left(1-P_{FA}^{(n)}[j+n-1]\right)$$
$$+Q\left(\frac{\gamma'_{rr}[j+n]-\mu_{rr,1}[j+n]}{\sigma_{rr,1}[j+n]}\right)P_{FA}^{(n)}[j+n-1]. \quad (117)$$

The average probability of error $(P_e)$ at the destination nanomachine corresponding to the transmission by the source nanomachine in slots 1 to $k$ for multi-hop molecular communication follows as

$$P_e = \frac{1}{k}\sum_{j=1}^{k}\left[\beta\left(1-P_D^d[j+N]\right)+(1-\beta)P_{FA}^d[j+N]\right]. \quad (118)$$

The mutual information $I(X[j], Y[j+N])$ between $X[j]$ and $Y[j+N]$ in multi-hop link can be expressed as

$$I(X[j], Y[j+N])$$
$$= \sum_{x[j]\in\{0,1\}}\sum_{y[j+N]\in\{0,1\}} \Pr(y[j+N]|x[j])\Pr(x[j])$$
$$\times \log_2 \frac{\Pr(y[j+N]|x[j])}{\sum_{x[j]\in\{0,1\}} \Pr(y[j+N]|x[j])\Pr(x[j])}, \quad (119)$$

where $\Pr(x[j]=1)=\beta$, $\Pr(x[j]=0)=1-\beta$, and $\Pr(y[j+N]\in\{0,1\}|x[j]\in\{0,1\})$ can be determined as

$$\Pr(y[j+N]=0|x[j]=0) = 1 - P_{FA}^d[j+N],$$
$$\Pr(y[j+N]=1|x[j]=0) = P_{FA}^d[j+N],$$
$$\Pr(y[j+N]=0|x[j]=1) = 1 - P_D^d[j+N],$$
$$\Pr(y[j+N]=1|x[j]=1) = P_D^d[j+N].$$

The capacity $C[k]$ of the multi-hop channel, when $k$ approaches $\infty$, can be obtained by maximizing the mutual information $I(X[j], Y[j+N])$ obtained corresponding to the transmission by the source-nanomachine in slots 1 to $k$ as

$$C[k] = \max_{\beta} \frac{1}{(k+N)}\sum_{j=1}^{k} I(X[j], Y[j+N]) \text{ bits/slot}. \quad (120)$$

## V. SIMULATION RESULTS

For simulation purposes, the distance between the source and destination nanomachines is set as $d_{sd} = 30$ $\mu$m, whereas the cooperative nanomachines in dual and multi-hop systems are located at distances $d_{sr} = d_{rd} = 15$ $\mu$m and $d_{sr_1} = d_{r_1r_2} = d_{r_2d} = 10$ $\mu$m respectively. The other parameters are diffusion coefficient $D = 2.2\times 10^{-11}$ m$^2$/s, degradation parameter $\alpha = 0.2$, slot duration $\tau \in \{2, 2.5, 3\}$s, prior probability $\beta = 0.5$, and drift velocity $v \in \{1.5\times 10^{-5}, 7\times 10^{-6}\}$ m/s. The MSI at each receiving node is modeled as a Gaussian distributed RV with mean $\mu_o = 20$ and variance $\sigma_o^2 = 20$. Similar to [20], the results are computed for a total of $K = 30$ slots with $\mathcal{Q}_0[j] = \mathcal{Q}_0, \mathcal{Q}_1[j+1] = \mathcal{Q}_1, \mathcal{Q}_2[j+2] = \mathcal{Q}_2, \forall j$.

Figs. 5-7 demonstrate the detection performance at the destination nanomachine in the diffusion based direct, dual-hop, and multi-hop molecular communication systems under various scenarios. For simulation purposes, the probabilities of detection and false alarm at the cooperative nanomachine $R$ in dual-hop communication as well as at the cooperative nanomachine $R_1$ in multi-hop communication are fixed as $P_D^{(1)}[j] = 0.99$ and $P_{FA}^{(1)}[j] = 0.01$ respectively. Fig. 5 shows the probability of detection versus probability of false alarm at the destination for a varying number of molecules at each transmitting nanomachine. One can observe that an increase in the number of molecules emitted by the nanomachines results

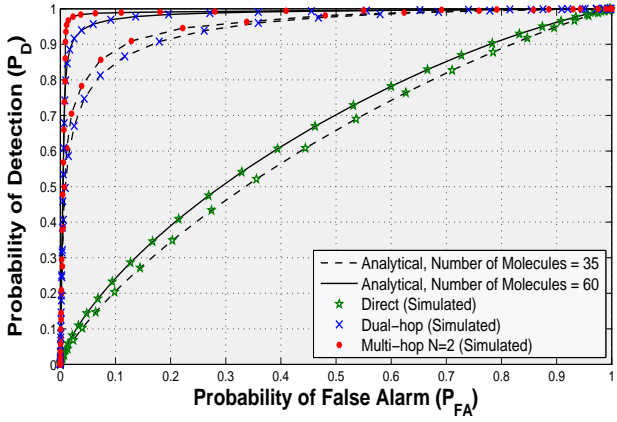

Fig. 5. Detection performance of direct, dual-hop, and multi-hop ($N = 2$) diffusion based molecular systems for a varying number of molecules with $D = 2.2 \times 10^{-11}$ m$^2$/s, $\tau = 2.5$ s, and $v = 7 \times 10^{-6}$ $\mu$m/s.

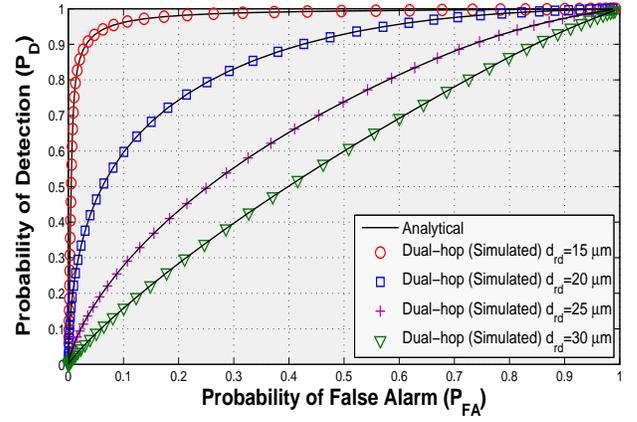

Fig. 7. Detection performance of a diffusion based dual-hop molecular communication system for various relay-destination distances with $\mathcal{Q}_0 = \mathcal{Q}_1 = 100$, $D = 2.2 \times 10^{-11}$ m$^2$/s, $\tau = 2.0$ s, and $v = 7 \times 10^{-6}$ $\mu$m/s.

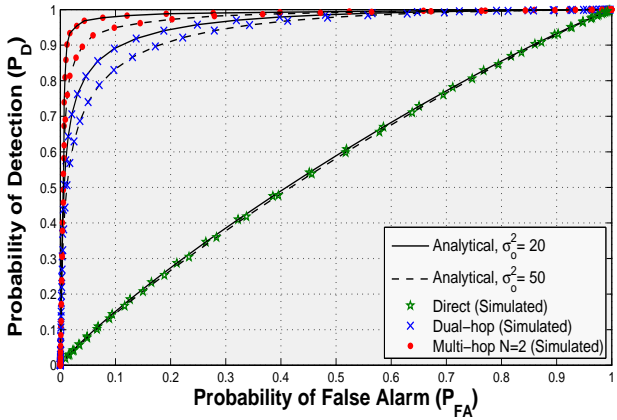

Fig. 6. Detection performance of direct, dual-hop, and multi-hop ($N = 2$) diffusion based molecular systems for a varying noise variance $\sigma_o^2$ with $\mu_o = 20$, $\mathcal{Q}_0 = \mathcal{Q}_1 = \mathcal{Q}_2 = 60$, $D = 2.2 \times 10^{-11}$ m$^2$/s, $\tau = 2.0$ s, and $v = 7 \times 10^{-6}$ $\mu$m/s.

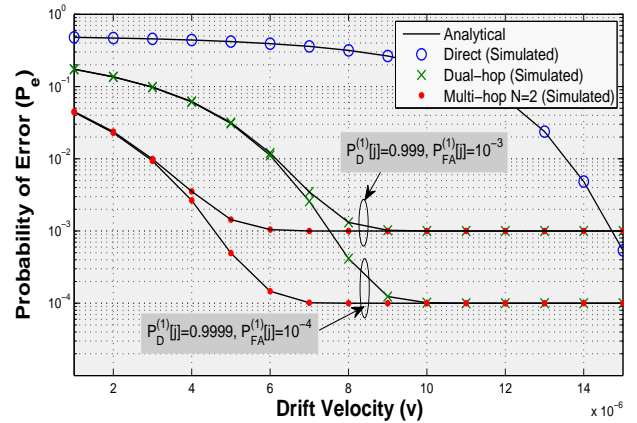

Fig. 8. Error rate performance of the direct, dual-hop, and multi-hop ($N = 2$) diffusion based molecular systems for a varying drift velocity $v$ with $\mu_o = \sigma_o^2 = 20$, $\mathcal{Q}_0 = \mathcal{Q}_1 = \mathcal{Q}_2 = 150$, $D = 2.2 \times 10^{-11}$ m$^2$/s, and $\tau = 2.5$ s.

in a higher probability of detection at the destination for a fixed value of probability of false alarm. It can also be clearly seen from Figs. 5-6 that the detection performance at the destination nanomachine significantly improves as the number of cooperative nanomachines $N$ increases between the source and destination nanomachines. One can also observe in Fig. 5 that for a fixed value of the probability of false alarm 0.1 and number of molecules 60, the probability of detection increases from 0.24 to 0.99 as the number of cooperative nanomachines $N$ increases from 0 to 2, where $N = 0$ represents direct communication. Similar improvement can also be seen in Fig. 6 for different values of noise variance $\sigma_o^2$.

Fig. 7 shows the probability of detection versus probability of false alarm at the destination nanomachine for various distances $d_{rd}$ between the cooperative and destination nanomachines in a dual-hop based molecular communication system. One can observe that the distance $d_{rd}$ between cooperative nanomachine $R$ and destination nanomachine has a significant impact on the detection performance at the destination. The probability of detection decreases significantly from approximately 0.96 to 0.16 at probability of false alarm 0.1 as the distance $d_{rd}$ increases from 15 $\mu$m to 30 $\mu$m.

Fig. 8 shows the probability of error $P_e$ versus the drift velocity $v$ for a fixed number of molecules $\mathcal{Q}_0 = \mathcal{Q}_1 = \mathcal{Q}_2 = 150$. First, it can be observed from Fig. 8 that the analytical $P_e$ values obtained using (23), (63), and (118) for direct, dual and multi-hop systems coincide with those obtained from simulations, thus validating the derived analytical results. One can also observe that the end-to-end performance of the system is significantly enhanced by cooperative nanomachines in comparison to the direct source-destination only communication scenario with low drift velocity. Moreover, similar to the detection performance, the end-to-end probability of error also decreases progressively as the number of cooperative nanomachines $N$ increases. Further, the dual-hop and multi-hop molecular communication systems experience a floor at high drift velocity, whereas the performance of direct communication system improves with increasing drift velocity.

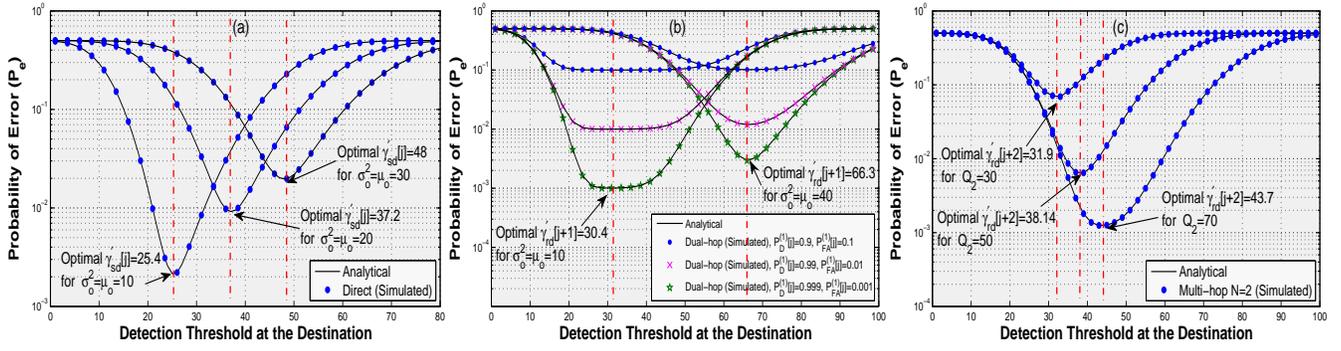

Fig. 9. Error rate of diffusion based molecular systems versus detection threshold for various scenarios with (a) direct communication with $\mathcal{Q}_0 = 60$ (b) dual-hop communication with $\mathcal{Q}_1 = 60$ (c) multi-hop ($N=2$) communication with $\mu_o = \sigma_o^2 = 20$, $P_D^{(1)}[j] = 0.999$, and $P_{FA}^{(1)}[j] = 0.001$.

This is owing to the fact that the end-to-end performance of dual and multi-hop systems at high drift velocity is dominated by the weak source-cooperative nanomachine link with fixed $(P_D^{(1)}[j], P_{FA}^{(1)}[j])$.

Fig. 9 shows the error rate performance versus detection threshold at the destination nanomachine in various scenarios. For this simulation, the diffusion coefficient, the slot duration and the drift velocity are set as, $D = 2.2 \times 10^{-11}$ m$^2$/s, $\tau = 3$ s, and $v = 1.5 \times 10^{-5}$ $\mu$m/s respectively. One can clearly observe that the optimal threshold obtained using (12), (47) and (78) for direct, dual-hop and multi-hop systems respectively, is able to achieve the minimum probability of error at the destination nanomachine. It can also be seen that the optimal values of the threshold at the destination depend on several parameters. The optimal value of the threshold for direct communication and dual-hop scenarios in Figs. 9(a)-(b) increases from 25.4 to 48 and 30.4 to 66.3 as the MSI increases from $\mu_o = \sigma_o^2 = 10$ to $\mu_o = \sigma_o^2 = 30$ and $\mu_o = \sigma_o^2 = 10$ to $\mu_o = \sigma_o^2 = 40$ respectively. A similar observation can also be made for multi-hop communication in Fig. 9(c) where the optimal threshold at the destination nanomachine increases from 31.9 to 43.7 as the number of molecules $\mathcal{Q}_2$ increases from 30 to 70.

Fig. 10 shows the capacity of diffusion based direct, dual-hop, and multi-hop molecular communication systems, where the maximum mutual information is achieved for equiprobable information symbols, i.e., $\beta = 0.5$. As depicted in Fig. 10(a), the capacity of molecular communication systems significantly reduces as the variance ($\sigma_o^2$) of MSI increases. Further, one can also observe that the capacity of the dual-hop and multi-hop systems depends highly on the detection performance of the intermediate cooperative nanomachines. As the detection performance $(P_D^{(1)}[j], P_{FA}^{(1)}[j])$ at $R$ in dual-hop and $R_1$ in multi-hop systems increases from $(0.9, 0.1)$ to $(0.99, 0.01)$, a significant capacity gain can be achieved at high MSI in comparison to the direct communication. However, at low MSI, the direct communication system achieves high capacity values in comparison with dual and multi-hop systems. This is due to the fixed probabilities of detection and false alarm at the intermediate cooperative nanomachine as well as the scaling factor $\frac{1}{k+N}$ associated with the effective capacity of the cooperative transmission. One can also notice that the multi-hop system achieves lower capacity values in comparison to the dual-hop system. This is due to the transmission between $R_1$ and $R_2$ in the multi-hop system that significantly degrades the detection performance at the destination nanomachine. On the other hand, the capacity of the molecular system without cooperating nanomachines increases significantly with an increase in the slot duration ($\tau$), as is clearly shown in Fig. 10(b). However, the capacity values obtained in dual and multi-hop systems are constant with increasing slot duration and are equal to the ones obtained for $P_D^d[j+1] = P_D^d[j+2] = P_D^{(1)}[j]$ and $P_{FA}^d[j+1] = P_{FA}^d[j+2] = P_{FA}^{(1)}[j]$. This is due to the fact the detection performance at the destination nanomachine is limited by the performance metrics $(P_D^{(1)}[j], P_{FA}^{(1)}[j])$ of the cooperative nanomachine $R$ in dual-hop and $R_1$ in multi-hop systems.

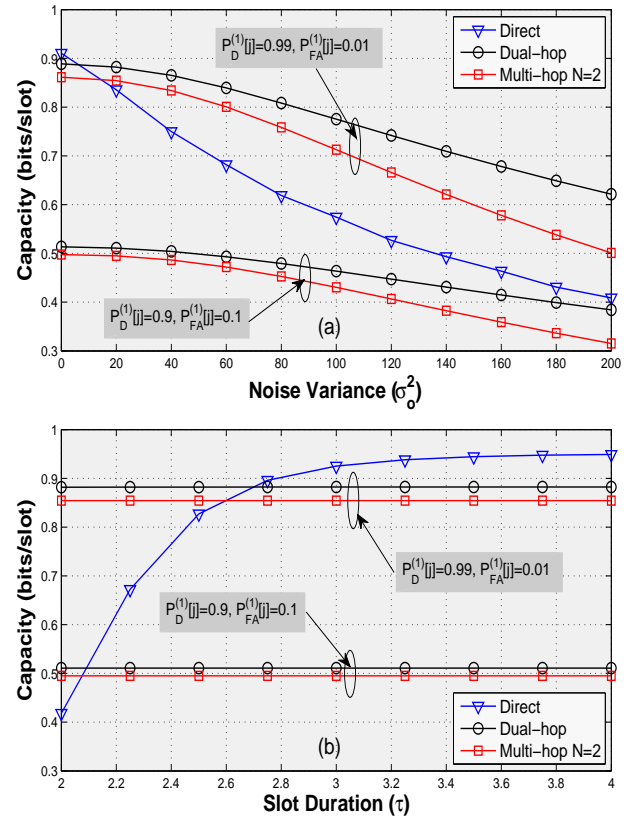

Fig. 10. Capacity of diffusion based direct, dual-hop and multi-hop molecular communication systems with $\mathcal{Q}_0 = \mathcal{Q}_1 = \mathcal{Q}_2 = 60$, $D = 2.2 \times 10^{-11}$ m$^2$/s, $v = 1.5 \times 10^{-5}$ $\mu$m/s, $\mu_o = 20$, and (a) increasing noise variance ($\sigma_o^2$) with $\tau = 2.5$ s (b) slot duration ($\tau$) with $\sigma_o^2 = 20$.

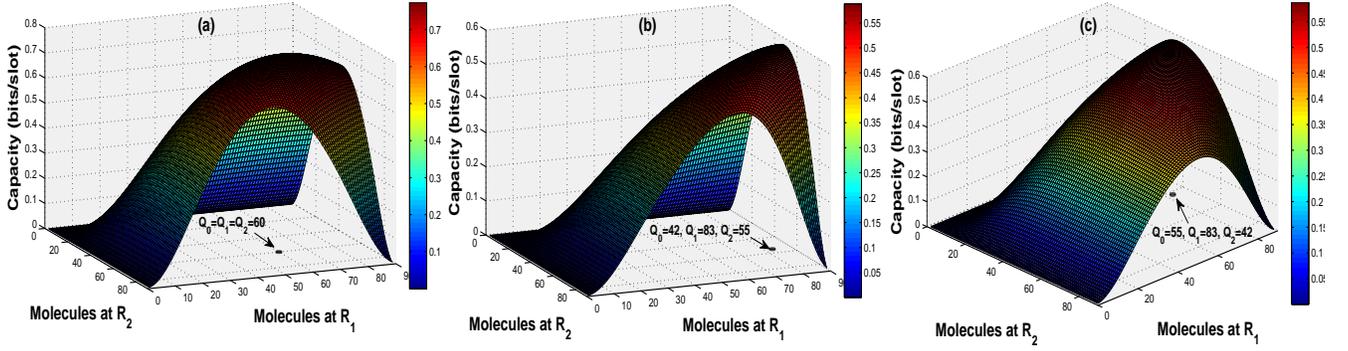

Fig. 11. Capacity of diffusion based multi-hop ($N = 2$) molecular system for different molecules allocation at each transmitting node with total molecule budget $\mathcal{Q}_0 + \mathcal{Q}_1 + \mathcal{Q}_2 = 180$, $\mu_o = 20, \sigma_o^2 = 10$, $D = 2.2 \times 10^{-11}$ m$^2$/s, $\tau = 2$ s, $v = 7 \times 10^{-6}$ $\mu$m/s, and distances (a) $d_{sr_1} = d_{r_1 r_2} = d_{r_2 d} = 10$ $\mu$m (b) $d_{sr_1} = 5$ $\mu$m, $d_{r_1 r_2} = 15$ $\mu$m, $d_{r_2 d} = 10$ $\mu$m (c) $d_{sr_1} = 10$ $\mu$m, $d_{r_1 r_2} = 15$ $\mu$m, $d_{r_2 d} = 5$ $\mu$m.

Fig. 11 finally demonstrates the impact of number of molecules allocated to each transmitting nanomachine considering a fixed molecule budget $\mathcal{Q}_0 + \mathcal{Q}_1 + \mathcal{Q}_2 = \mathcal{Q} = 180$ and various distance configurations in a multi-hop molecular communication system. Firstly, One can observe that allocating equal number of molecules, $\mathcal{Q}_0 = \mathcal{Q}_1 = \mathcal{Q}_2 = \frac{\mathcal{Q}}{3} = 60$, at the source and cooperative nanomachines $R_1$ and $R_2$ is only optimal for the scenarios when the distances between all communicating nanomachines are equal. For the scenarios when $d_{r_1 r_2} > d_{r_2 d} > d_{sr_1}$ and $d_{r_1 r_2} > d_{sr_1} > d_{r_2 d}$ as shown in Fig. 11(b) and Fig. 11(c) respectively, the system capacity can be significantly enhanced by allocating a large fraction of the molecule budget $\mathcal{Q}$ to cooperative nanomachine $R_1$ in comparison to the source and cooperative nanomachine $R_2$.

## VI. CONCLUSION

This work analyzed the performance of diffusion based direct, dual-hop, and multi-hop molecular communication systems in the presence of ISI, MSI, and counting errors along with Brownian motion and drift. Analytical expressions were derived for the optimal test statistics and optimal thresholds at the cooperative and destination nanomachines, together with the resulting probabilities of detection, false alarm as well as the end-to-end probability of error for each scenario. In addition, the capacity expressions were also derived. It was seen that the detection performance at the destination nanomachine can be significantly enhanced by incorporating cooperative nanomachines between the source and destination nanomachines. Further, this work also demonstrated that the end-to-end performance of dual and multi-hop molecular systems is dominated by the performance of the weaker link. Finally, future studies can focus on the impact of mobile nanomachines as well as the optimization of transmitted molecules in each time-slot on the performance of relay-assisted diffusive mobile molecular communication.

## APPENDIX A
### DERIVATION OF EXPRESSION (54)

Substituting the expressions for $p(R_{rd}[j+1]|\widehat{x}[j] = 1)$ and $p(R_{rd}[j+1]|\widehat{x}[j] = 0)$ in (51) and cross multiplying with the threshold, the expression can be simplified as shown in (121)-(123). Now, taking the logarithm of both sides of (123), the resulting expression can be simplified to yield the expression given in (54).

## APPENDIX B
### DERIVATION OF EXPRESSION (90)

Substituting the expressions given below for $p(R_{rd}[j+N]|\xi_l), l = 1, 3, \cdots, 2^N - 1$ and $p(R_{rd}[j+N]|\xi_l), l = 0, 2, \cdots, 2^N - 2$ in (83)

$$p(R_{rd}[j+N]|\xi_l) = \begin{cases} \frac{1}{\sqrt{2\pi\sigma_{rd,1}^2[j+N]}} \\ \times \exp\left(-\frac{(R_{rd}[j+N]-\mu_{rd,1}[j+N])^2}{2\sigma_{rd,1}^2[j+N]}\right), l=1,\cdots,2^N-1 \\ \frac{1}{\sqrt{2\pi\sigma_{rd,0}^2[j+N]}} \\ \times \exp\left(-\frac{(R_{rd}[j+N]-\mu_{rd,0}[j+N])^2}{2\sigma_{rd,0}^2[j+N]}\right), l=2,\cdots,2^N-2. \end{cases}$$
(124)

and cross multiplying with the threshold $\frac{1-\beta}{\beta}$, the expression can be simplified as shown in (125)-(127). Now, taking the logarithm of both sides of (127), the resulting expression can be simplified to yield the expression given in (90). where $\beta_{rd}$ is defined as,

$$\beta_{rd} = \left[\frac{\sum_{l=0,2,\cdots,2^N-2}[(1-\beta)\Pr(\xi_l|\mathcal{H}_0) - \beta\Pr(\xi_l|\mathcal{H}_1)]}{\sum_{l=1,3,\cdots,2^N-1}[\beta\Pr(\xi_l|\mathcal{H}_1) - (1-\beta)\Pr(\xi_l|\mathcal{H}_0)]}\right], \quad (128)$$

where the probability of the system being in state $\xi_l$ under $\mathcal{H}_0$ and $\mathcal{H}_1$ can be expressed as,

$$\Pr(\xi_l|\mathcal{H}_0) = \prod_{n \in \Psi_l} P_{FA}^{(n)}[j+n-1] \prod_{n \in \overline{\Psi}_l} \left(1 - P_{FA}^{(n)}[j+n-1]\right),$$

$$\Pr(\xi_l|\mathcal{H}_1) = \prod_{n \in \Psi_l} P_D^{(n)}[j+n-1] \prod_{n \in \overline{\Psi}_l} \left(1 - P_D^{(n)}[j+n-1]\right).$$

$$\beta \left[ \frac{P_D^{(1)}[j]}{\sqrt{2\sigma_{rd,1}^2[j+1]}} \exp\left(-\frac{(R_{rd}[j+1]-\mu_{rd,1}[j+1])^2}{2\sigma_{rd,1}^2[j+1]}\right) + \frac{(1-P_D^{(1)}[j])}{\sqrt{2\sigma_{rd,0}^2[j+1]}} \exp\left(-\frac{(R_{rd}[j+1]-\mu_{rd,0}[j+1])^2}{2\sigma_{rd,0}^2[j+1]}\right) \right]$$

$$\underset{\mathcal{H}_0}{\overset{\mathcal{H}_1}{\gtreqless}} (1-\beta) \left[ \frac{P_{FA}^{(1)}[j]}{\sqrt{2\sigma_{rd,1}^2[j+1]}} \exp\left(-\frac{(R_{rd}[j+1]-\mu_{rd,1}[j+1])^2}{2\sigma_{rd,1}^2[j+1]}\right) + \frac{(1-P_{FA}^{(1)}[j])}{\sqrt{2\sigma_{rd,0}^2[j+1]}} \left(-\frac{(R_{rd}[j+1]-\mu_{rd,0}[j+1])^2}{2\sigma_{rd,0}^2[j+1]}\right) \right], \quad (121)$$

$$\frac{(\beta P_D^{(1)}[j] - (1-\beta) P_{FA}^{(1)}[j])}{\sqrt{\sigma_{rd,1}^2[j+1]}} \exp\left(-\frac{(R_{rd}[j+1]-\mu_{rd,1}[j+1])^2}{2\sigma_{rd,1}^2[j+1]}\right)$$

$$\underset{\mathcal{H}_0}{\overset{\mathcal{H}_1}{\gtreqless}} \frac{[(1-\beta)(1-P_{FA}^{(1)}[j]) - \beta(1-P_D^{(1)}[j])]}{\sqrt{\sigma_{rd,0}^2[j+1]}} \exp\left(-\frac{(R_{rd}[j+1]-\mu_{rd,0}[j+1])^2}{2\sigma_{rd,0}^2[j+1]}\right), \quad (122)$$

$$\exp\left(\frac{(R_{rd}[j+1]-\mu_{rd,0}[j+1])^2}{2\sigma_{rd,0}^2[j+1]} - \frac{(R_{rd}[j+1]-\mu_{rd,1}[j+1])^2}{2\sigma_{rd,1}^2[j+1]}\right) \underset{\mathcal{H}_0}{\overset{\mathcal{H}_1}{\gtreqless}} \sqrt{\frac{\sigma_{rd,1}^2[j+1]}{\sigma_{rd,0}^2[j+1]}} \left[ \frac{(1-\beta)(1-P_{FA}^{(1)}[j]) - \beta(1-P_D^{(1)}[j])}{\beta P_D^{(1)}[j] - (1-\beta) P_{FA}^{(1)}[j]} \right]. \quad (123)$$

$$\beta \left[ \sum_{l=0,2,\cdots,2^N-2} \frac{\Pr(\xi_l|\mathcal{H}_1)}{\sqrt{2\sigma_{rd,1}^2[j+N]}} \exp\left(-\frac{(R_{rd}[j+N]-\mu_{rd,1}[j+N])^2}{2\sigma_{rd,1}^2[j+N]}\right) \right.$$

$$\left. + \sum_{l=1,3,\cdots,2^N-1} \frac{\Pr(\xi_l|\mathcal{H}_1)}{\sqrt{2\sigma_{rd,0}^2[j+N]}} \exp\left(-\frac{(R_{rd}[j+N]-\mu_{rd,0}[j+N])^2}{2\sigma_{rd,0}^2[j+N]}\right) \right]$$

$$\underset{\mathcal{H}_0}{\overset{\mathcal{H}_1}{\gtreqless}} (1-\beta) \left[ \sum_{l=0,2,\cdots,2^N-2} \frac{\Pr(\xi_l|\mathcal{H}_0)}{\sqrt{2\sigma_{rd,1}^2[j+N]}} \exp\left(-\frac{(R_{rd}[j+N]-\mu_{rd,1}[j+N])^2}{2\sigma_{rd,1}^2[j+N]}\right) \right.$$

$$\left. + \sum_{l=1,3,\cdots,2^N-1} \frac{\Pr(\xi_l|\mathcal{H}_0)}{\sqrt{2\sigma_{rd,0}^2[j+N]}} \exp\left(-\frac{(R_{rd}[j+N]-\mu_{rd,0}[j+N])^2}{2\sigma_{rd,0}^2[j+N]}\right) \right], \quad (125)$$

Substituting now the above expressions in (128), the expression for $\beta_{rd}$ can be further simplified for different number of cooperative nanomachines as follows. For $N=2$ cooperative nanomachines, the expression for $\beta_{rd}$ can be solved considering the four possible states, i.e., $\xi_0 = [0,0], \xi_1 = [0,1], \xi_2 = [1,0]$ and $\xi_3 = [1,1]$ as $\beta_{rd} = \frac{N_2}{D_2}$, where $N_2$ and $D_2$ are given in (129) and (130) respectively. The expressions in (129) and (130) can be further simplified to yield the final expression for $\beta_{rd}$ for $N=2$ cooperative nanomachines as,

$$\beta_{rd} = \frac{(1-\beta)\left(1-P_{FA}^{(2)}[j+1]\right) - \beta\left(1-P_D^{(2)}[j+1]\right)}{\beta P_D^{(2)}[j+1] - (1-\beta) P_{FA}^{(2)}[j+1]}. \quad (131)$$

The expression for $\beta_{rd}$ considering the eight possible states for $N=3$ cooperative nanomachines can be obtained as $\beta_{rd} = \frac{N_3}{D_3}$, where $N_3$ and $D_3$ are given in (132) and (133) respectively. The expressions in (132) and (133) can be further simplified to yield the final expression for $\beta_{rd}$ for $N=3$ cooperative nanomachines as,

$$\beta_{rd} = \frac{(1-\beta)\left(1-P_{FA}^{(3)}[j+2]\right) - \beta\left(1-P_D^{(3)}[j+2]\right)}{\beta P_D^{(3)}[j+2] - (1-\beta) P_{FA}^{(3)}[j+2]}. \quad (134)$$

On the similar lines, the expression for $\beta_{rd}$ considering $2^N$ possible states due to the presence of $N$ cooperative nanomachines can be similarly obtained as,

$$\beta_{rd} = \frac{(1-\beta)\left(1-P_{FA}^{(N)}[j+N-1]\right) - \beta\left(1-P_D^{(N)}[j+N-1]\right)}{\beta P_D^{(N)}[j+N-1] - (1-\beta) P_{FA}^{(N)}[j+N-1]}. \quad (135)$$

$$\frac{1}{\sqrt{\sigma_{rd,1}^2[j+N]}} \exp\left(-\frac{(R_{rd}[j+N]-\mu_{rd,1}[j+N])^2}{2\sigma_{rd,1}^2[j+N]}\right) \sum_{l=1,3,\cdots,2^N-1} [\beta \Pr(\xi_l|\mathcal{H}_1) - (1-\beta)\Pr(\xi_l|\mathcal{H}_0)]$$

$$\mathop{\gtreqless}\limits_{\mathcal{H}_0}^{\mathcal{H}_1} \frac{1}{\sqrt{\sigma_{rd,0}^2[j+N]}} \exp\left(-\frac{(R_{rd}[j+N]-\mu_{rd,0}[j+N])^2}{2\sigma_{rd,0}^2[j+N]}\right) \sum_{l=0,2,\cdots,2^N-2} [(1-\beta)\Pr(\xi_l|\mathcal{H}_0) - \beta\Pr(\xi_l|\mathcal{H}_1)], \quad (126)$$

$$\exp\left(\frac{(R_{rd}[j+N]-\mu_{rd,0}[j+N])^2}{2\sigma_{rd,0}^2[j+N]} - \frac{(R_{rd}[j+N]-\mu_{rd,1}[j+N])^2}{2\sigma_{rd,1}^2[j+N]}\right) \mathop{\gtreqless}\limits_{\mathcal{H}_0}^{\mathcal{H}_1} \sqrt{\frac{\sigma_{rd,1}^2[j+N]}{\sigma_{rd,0}^2[j+N]}} \beta_{rd}, \quad (127)$$

$$N_2 = (1-\beta)\left(1-P_{FA}^{(1)}[j]\right)\left(1-P_{FA}^{(2)}[j+1]\right) - \beta\left(1-P_D^{(1)}[j]\right)\left(1-P_D^{(2)}[j+1]\right)$$
$$+ (1-\beta)P_{FA}^{(1)}[j]\left(1-P_{FA}^{(2)}[j+1]\right) - \beta P_D^{(1)}[j]\left(1-P_D^{(2)}[j+1]\right), \quad (129)$$

$$D_2 = \beta\left(1-P_D^{(1)}[j]\right)P_D^{(2)}[j+1] - \left(1-\beta\right)(1-P_{FA}^{(1)}[j])P_{FA}^{(2)}[j+1] + \beta P_D^{(1)}[j]P_D^{(2)}[j+1] - (1-\beta)P_{FA}^{(1)}[j]P_{FA}^{(2)}[j+1]. \quad (130)$$

$$\begin{aligned}
N_3 =&(1-\beta)\left(1-P_{FA}^{(1)}[j]\right)\left(1-P_{FA}^{(2)}[j+1]\right)\left(1-P_{FA}^{(3)}[j+2]\right)-\beta\left(1-P_D^{(1)}[j]\right)\left(1-P_D^{(2)}[j+1]\right)\left(1-P_D^{(3)}[j+2]\right)\\
&+(1-\beta)\left(1-P_{FA}^{(1)}[j]\right)P_{FA}^{(2)}[j+1]\left(1-P_{FA}^{(3)}[j+2]\right)-\beta\left(1-P_D^{(1)}[j]\right)P_D^{(2)}[j+1]\left(1-P_D^{(3)}[j+2]\right)\\
&+(1-\beta)P_{FA}^{(1)}[j]\left(1-P_{FA}^{(2)}[j+1]\right)\left(1-P_{FA}^{(3)}[j+2]\right)-\beta P_D^{(1)}[j]\left(1-P_D^{(2)}[j+1]\right)\left(1-P_D^{(3)}[j+2]\right)\\
&+(1-\beta)P_{FA}^{(1)}[j]P_{FA}^{(2)}[j+1]\left(1-P_{FA}^{(3)}[j+2]\right)-\beta P_D^{(1)}[j]P_D^{(2)}[j+1]\left(1-P_D^{(3)}[j+2]\right),
\end{aligned} \quad (132)$$

$$\begin{aligned}
D_3 =&\beta\left(1-P_D^{(1)}[j]\right)\left(1-P_D^{(2)}[j+1]\right)P_D^{(3)}[j+2]-(1-\beta)\left(1-P_{FA}^{(1)}[j]\right)\left(1-P_{FA}^{(2)}[j+1]\right)P_{FA}^{(3)}[j+2]\\
&+\beta\left(1-P_D^{(1)}[j]\right)P_D^{(2)}[j+1]P_D^{(3)}[j+2]-(1-\beta)\left(1-P_{FA}^{(1)}[j]\right)P_{FA}^{(2)}[j+1]P_{FA}^{(3)}[j+2]\\
&+\beta P_D^{(1)}[j]\left(1-P_D^{(2)}[j+1]\right)P_D^{(3)}[j+2]-(1-\beta)P_{FA}^{(1)}[j]\left(1-P_{FA}^{(2)}[j+1]\right)P_{FA}^{(3)}[j+2]\\
&+\beta P_D^{(1)}[j]P_D^{(2)}[j+1]P_D^{(3)}[j+2]-(1-\beta)P_{FA}^{(1)}[j]P_{FA}^{(2)}[j+1]P_{FA}^{(3)}[j+2].
\end{aligned} \quad (133)$$